\documentstyle[11pt]{article} 
\def\d{{\rm d}}
\def\exp{{\rm exp}}

\def\bbbone{{\mathchoice {\rm 1\mskip-4mu {\rm l}} 
  {\rm 1\mskip-4mu {\rm l}} {\rm 1\mskip-4.5mu
  {\rm l}} {\rm 1\mskip-5mu {\rm l}}}}
\newcommand{\bbbr}{{\kern+,25em\sf{R}\kern-,78em\sf{I}
  \kern+,78em\kern-,25em}}
\newcommand{\bbbc}{{\sf{C}\kern-.46em\sf{I}\kern+.46em
  \kern-.25em}}
\newcommand{\bbbz}{{\kern+.25em\sf{Z}\kern-.78em\sf{Z}
  \kern+.78em\kern-.65em}}

\long\def\@makecaption#1#2{
   \vskip 10pt 
   \setbox\@tempboxa\hbox{\footnotesize #1: #2}
 \ifdim \wd\@tempboxa >\hsize \footnotesize #1: #2\par 
\else \hbox
 to\hsize{\hfil\box\@tempboxa\hfil}     \fi}

\begin{document} 
\vspace*{4 cm} 
\Large{\bf Quantization of Kinematics on Configuration
 Manifolds\footnote{Reviews in Mathematical Physics,
Vol. 13, No. 4 (2001), pp. 1--47.}
}\\
\normalsize \\[6ex]
\hspace*{3 cm} H.-D. Doebner\\ 
\hspace*{3 cm} Arnold Sommerfeld
Institut f\"ur Mathematische Physik\\ 
\hspace*{3 cm} Technische Universit\"at Clausthal\\
\hspace*{3 cm} Leibnizstr. 10 \\ 
 \hspace*{3 cm} D-38678 Clausthal (Fed. Rep. Germany)\\[3ex]
 \hspace*{3 cm} P. \v{S}\v{t}ov\'\i\v{c}ek and J. Tolar\\
\hspace*{3 cm} Doppler Institute\\ 
 \hspace*{3 cm} Faculty of Nuclear Sciences and
Physical Engineering\\ 
\hspace*{3 cm} Czech Technical University\\ 
\hspace*{3 cm} B\v{r}ehov\'a 7\\ 
\hspace*{3 cm} CZ-115 19 Prague 1 (Czech Republic)\\[3ex]
 \vspace{1cm} 

\begin{abstract} 
The review is devoted to topological global aspects
of quantal description. The treatment concentrates
on quantizations of kinematical observables --- generalized
positions and momenta. A broad class of quantum kinematics
is rigorously constructed for systems, the configuration
space of which is either a homogeneous space of a Lie group
or a connected smooth finite-dimensional manifold without
boundary. The class also includes systems in an external
gauge field for an Abelian or a compact gauge group.
Conditions for equivalence and irreducibility of 
generalized quantum kinematics are investigated with the aim
of classification of possible quantizations. Complete
classification theorems are given in two special cases.
It is attempted to motivate the global approach based
on a generalization of imprimitivity systems called
{\em quantum Borel kinematics}. These are classified by
means of global invariants --- quantum numbers of
topological origin. Selected examples are presented which
demonstrate the richness of applications of Borel
quantization. 
The review aims to provide an introductory
survey of the subject and to be sufficiently
selfcontained as well, 
so that it can serve as a standard reference
concerning Borel quantization for systems
admitting localization on differentiable manifolds.

\end{abstract}

\newpage
\tableofcontents
\pagebreak

\section{Introduction}

The successful development of quantum theory in this century
shows convincingly that it provides perhaps the most
universal language for the description of physical
phenomena. In quantum theory, as in any other physical
theory, two fundamental aspects can be distinguished: 
the mathematical formalism and the physical interpretation.

At the basis of the most common {\it mathematical formalism}
of quantum mechanics lies the notion of a complex separable
Hilbert space $\cal{H}$ of, in general, infinite dimension.
Normed vectors in $\cal{H}$ correspond to pure states of
a quantum system, whereas quantal observables are represented by self-adjoint operators in $\cal{H}$.

However, only the rules of a {\it physical interpretation}
enable one to use quantum theory for the description of
physical systems. The principal general rule is Born's
statistical interpretation of the wave function.
For each physical system, or at least for a certain class of
them, it is further necessary to specify which
operators in $\cal{H}$ are associated with physical observables measured by certain measuring devices. 
This means in particular that at least the operators of
kinematical observables (position and momentum), and
the dynamical evolution law of the system are to be specified.

An important tool for the derivation of quantum models
are quantization methods.
The primary aim of {\it quantization} of a given classical
system is to associate self-adjoint operators with classical
observables. As a rule, two main methods are used.
The first one is based on Bohr's {\it correspondence principle}:
the physical meaning of quantum operators is found
by looking at their classical counterparts.
In this way non-relativistic quantum mechanics was formulated
by quantization of classical Hamiltonian mechanics,
quantum theory of electromagnetic field by quantization
of the Maxwell theory, etc. \cite{Dirac}.

The correspondence principle can, of course, be the leading
rule for quantization, if the observables already existed in
a classical form. What should be done in the case of
quantum observables without a classical analogue like the spin?
Here the second method is often applicable, which uses 
{\it invariance principles} connected with the symmetries 
of the system. By Noether's theorem the operators
corresponding to conserved quantities can be found as
generators of some projective representation of the symmetry
group in ${\cal H}$.
As a far reaching application of this approach let us
mention the relativistic quantum theory of elementary
particles based on the irreducible unitary representations
of the Poincar\'{e} group.

Both methods were used from the very first days of quantum
theory, always taking into account specific physical
properties of the systems considered. The first method
usually appears in non-relativistic quantum mechanics as
{\it canonical quantization} \cite{Dirac}, for systems with
the Euclidean configuration space $\bbbr^n$. 
The position coordinates $q_j$ and the canonically conjugate
momenta $p_k$ are quantized
into self-adjoint position and momentum operators
$Q_j$, $P_k$ (in a separable Hilbert space ${\cal H}$),
satisfying canonical commutation relations. This was
originally discovered and mathematically formulated
independently by W. Heisenberg and E. Schr\"{o}dinger in
1925--26. The uniqueness of the
mathematical formulation up to unitary equivalence was then
guaranteed by the Stone-von Neumann Theorem.

Quantum mechanics on $\bbbr^n$ became very soon a successful
theory which has been able to correctly describe
experimental findings in vast areas of quantum physics.
However, in some cases it was necessary 
to look for a formulation of quantum mechanics when 
the configuration space of a system was not Euclidean
\cite{28}. For instance, in connection with the studies 
of rotational spectra of molecules and of deformed nuclei,
{\it quantum rotators} were introduced as fundamental
quantum models with configuration spaces $S^1$ (the circle),
$S^2$ (the 2-sphere) and $SO(3)$ (the rotation group).
The textbook treatment of spinning top models 
(quantum mechanics of angular momentum) 
presents a successful
application of the approach via invariance principles.

There were also attempts to enforce canonical quantization
in cases where global Cartesian coordinates do not exist on
the configuration manifold $M$. A formal quantization
of generalized coordinates $q_j$ and conjugate momenta
$p_k$ was suggested \cite{28} on a manifold $M$ with the
 Riemann structure  (metric tensor $g_{jk}$ with
 determinant $g>0$):
$$
Q_j = q_j, \quad P_k = -i\hbar{\partial \over \partial q_k} -
{i\hbar \over 4} {\partial \over \partial q_k}(\ln g)\,.
\eqno{(1.1)}
$$
Note that the additional term in $P_k$'s makes them
symmetric operators in ${\cal H} = L^2 (M, {\rm d}\mu)$
with respect to the Riemann measure 
${\rm d}\mu = \sqrt g\, {\rm d}^{n}q$ on $M$. 
The main difficulty encountered
here is that operators (1.1) are not globally defined
since, in general, $q_j$ are only {\it local coordinates}.
It is therefore desirable to invent quantization methods
which employ {\it global geometric objects}.

On several occasions the formalism of quantum mechanics in
 connection with 
non--trivial topology of the configuration manifold 
lead to new non--classical effects.
A deep and in its time not completely understood and
recognized accomplishment in this direction was Dirac's
famous investigation \cite{5} of a quantum charged particle
(charge $e$) in the external magnetic field of 
a point--like magnetic monopole (magnetic charge $g$).
If the singular {\it Dirac monopole} is placed at the
origin of a Cartesian coordinate system in $\bbbr^3$,
one deals in fact with quantum mechanics on a topologically
non-trivial effective configuration manifold 
$\bbbr^{3} \backslash \{(0,0,0)\}$ 
(the three-dimensional Euclidean space with the
origin excluded). Here the formalism of quantum mechanics
in connection with non-trivial topology of the configuration
manifold leads to an unexpected topological quantum effect
originating from a peculiar behaviour of the phase of
a wave function:
Dirac discovered that a quantal description exists only under the condition that the dimensionless quantity $eg/2\pi\hbar$ is an integer.

Another phenomenon of this kind was noticed in 1959 by 
Y. Aharonov and D. Bohm \cite{1}. The origin of the 
Aharonov--Bohm effect can be traced to a shift of the phase
of wave function due to an external magnetic flux imposed on
a charged particle. Here the effective Aharonov--Bohm
configuration space is $\bbbr^{3} \backslash \bbbr$, 
the three--dimensional Euclidean space with a straight line
excluded.
In both mentioned cases the topologies of
the configuration spaces differ from the trivial topology
of the Euclidean space and play decisive r\^{o}le in
quantum theory.

These remarks about the early history of quantum mechanics
clearly point to the need for a systematic development of
{\it global quantization methods}.
For systems with sufficiently {\it symmetric} configuration
or phase spaces, two modern approaches in the theory of
group representations can be applied:
\begin{description}
\item {1.}  Mackey's quantization on homogeneous
 configuration  manifolds  $M = G/H$ \cite{7, 11, 26}.
 Essentially, it is equivalent to the construction of 
{\em systems of imprimitivity} for a
(locally compact, separable) group $G$, based on $M = G/H$.
\item {2.}  The method of {\it coadjoint orbits} which play
 the  role of homogeneous phase spaces \cite{Ko,23}.
\end{description}
In the case of configuration or phase manifolds 
{\it without} geometric symmetries, two programs of 
global quantization were suggested:
\begin{description}
\item {3.} {\it Borel quantization} on configuration
 manifolds \cite{4,8} which extends the notion of Schr\"odinger
 systems \cite{12}.
\item {4.}  {\it Geometric quantization} on symplectic phase
manifolds \cite{Ko,23}.
\end{description}
These methods have been elaborated to differing degrees of
sophistication and have, in general, different classes of
classical systems as their domains of applicability. 
{\it Borel quantization} is built on configuration spaces
 and reflects the topology of $M$. 
For physical applications it
is important that it yields both important classification
theorems and explicit relations for quantization of
kinematical observables. Like canonical quantization,
it is a two step procedure. In a first step the kinematics,
i.e. position and momentum observables on $M$, is quantized.
The time dependence is introduced in a second step with
a quantum analogue of a second order Riemannian dynamics
on $M$ \cite{4}. In its most general form it leads to 
Doebner--Goldin non--linear Schr\"odinger equations
 \cite{DG}.

Concerning other quantization methods respecting global
properties of confi\-gu\-ration or phase spaces we should
especially mention:
\begin{description}
\item {5.} The {\it Feynman path integral} method
 (it was used, e.g., in \cite{13} for $M$ = SO(3) and 
in \cite{9} for configuration spaces of identical particles).
\item {6.} {\it Quantization by deformation} of classical
 mechanics \cite{BFFLS}.
\item {7.} {\it Dirac quantization} of systems with
 constraints in phase space \cite{Dir}.
\end{description}

This review article is devoted to the mathematical
exposition of {\it quantum Borel kinematics}.
This method yields quantizations of kinematics
for systems admitting localization on connected
smooth finite-dimensional configuration manifolds without
boundary. We restrict our consideration exclusively to
paracompact manifolds which (by Whitney's embedding theorem)
 can be regarded as submanifolds of $\bbbr^n$.

In Chap. 2, the Hilbert space formalism of quantum
 mechanics, Wigner's Theorem on symmetry transformations,
 and the notion of Mackey's system of imprimitivity are
 briefly surveyed. The notion of quantum Borel kinematics 
is introduced in Chap. 3. In Chap. 4, a family of quantum
 Borel kinematics is constructed. This geometrical
 construction of quantum kinematics (Sect. 4.2) is based 
on the notion of a generalized
 system of imprimitivity for the family of one-parameter
 groups of diffeomorphisms (Sect. 3.1). It represents a
 generalization of quantum Borel kinematics of
 Ref. \cite{4}, especially in admitting an external gauge
 field with an arbitrary Abelian or compact structure group
 $G$. Thus the construction involves associated 
$\bbbc^{r}$-bundles with
{\it finite-dimensional fibres}~$\bbbc^r$.
Chap. 4 is also devoted to questions of
 unitary equivalence and irreducibility (Sects. 4.5 and 4.6)
 of this class of quantum kinematics. 
Important special case of the vanishing external field
is treated in Sect. 4.7.
The classification of quantum Borel kinematics cannot
be considered to be complete. In Chap. 5 theorems are stated
 which fully characterize them as well as two cases
of complete classifications --- elementary quantum Borel
kinematics (Sect. 5.5) and quantum Borel $r$--kinematics
of type 0 (Sect. 5.6).  In these cases it is shown
 that the first and the second singular homology groups of
 the configuration manifold $M$ are involved and
provide the necessary
 topological tools for classification of quantizations.

We have payed particular attention to a selection of proper
examples which complement each chapter and demonstrate the
richness of possible applications. From these
examples, we mention a new derivation of the Dirac
 quantization condition from rotational symmetry \cite{15}
 (Ex. 2.4), a topological description of the Aharonov--Bohm
 effect (Ex. 5.2), classification of elementary quantum
Borel kinematics on arbitrary
two--dimensional compact orientable manifolds (Ex. 5.3) as
well as in the real projective space ---
topologically non--trivial part of the configuration space
of the system of two identical particles
\cite{6} (Ex. 5.4). 

\section{Mackey's system of imprimitivity}
\subsection{The formalism of quantum mechanics} 
In quantum mechanics, a separable Hilbert space 
${\cal H}$ is associated with a quantum system 
we are going to describe. {\it States} of
the system are represented by von Neumann's statistical
(density) operators --- bounded self--adjoint positive
operators in ${\cal H}$ with unit trace. 
The set of states ${\cal W}$
introduced in this way is convex; 
its extremal points are called {\it pure states}. 
The pure states are just the projectors on
one--dimensional subspaces of ${\cal H}$.

We assume\footnote{We are going to use such spectral
measures for position measurements 
in configuration space and assume tacitly that they
are ideal and the state after 
the measurement can be described by a projection
$E(S)\psi$ of the original state $\psi$. There are
several options for a description of a non--ideal 
localization --- e.g. with the use of positive
operator--valued measures \cite{Davies}, Ch. 3.}
with \cite{29} that, to a measurement on the system, taking
values in a set $X$ endowed with a $\sigma$-algebra 
${\cal B}(X)$ of measurable subsets, there corresponds 
a projection--valued measure $F$ on $X$. 
To any measurable set $S\in {\cal B}(X)$, a projector
$F(S)$ is related such that $F(X)=\bbbone$ and 
$F(\bigcup_{i=1}^{\infty}S_i)=\sum_{i=1}^{\infty}F(S_i)$,
 provided $S_i \cap S_j = \emptyset$ for $i\neq j$. 
If the system is in a state
$U\in {\cal W}$, then the formula 
$$p_U (S) = {\rm Tr}(UF(S))$$
gives the probability that the result of the measurement belongs to the set $S\in{\cal B}(X)$. The map
 $p_U\colon{\cal B}(X)\rightarrow \bbbr$ is evidently 
a probability measure on $X$.

From the above considerations it is clear that in the quantal
formalism crucial role is played by an \em orthocomplemented
lattice of projectors \em onto subspaces of the Hilbert space
${\cal H}$ \cite{29}. This lattice will be denoted by ${\cal
L}$; partial ordering of ${\cal L}$ is defined as follows:
$F_1\leq F_2$ if and only if $F_{1}F_2 =F_1$; the complement:
$F^\bot = \bbbone -F$. Clearly, $F_1 \leq F_2$ if and only
if the corresponding subspaces are in inclusion; $F^\bot$
projects on the orthogonal complement. Important properties of
the lattice of projectors are: 
\begin{enumerate} 
   \item[$(\imath)$] For any
countable set $F_1, F_2, \dots$ of elements from ${\cal L}$
there exist $\bigvee_{n}F_n$ and $\bigwedge_{n}F_n$ in ${\cal L}$;
\footnote{
The element  $A = \bigvee_{n}F_n$ is defined by the
following properties: 1)$A \geq F_n$ for all $n$; 
2) if $B$ is any element of  ${\cal L}$ such that 
$F_{n} \leq B$ for all $n$, then $A \leq B$. 
In an analogous fashion,
the element $C = \bigwedge_{n}F_n$ is defined by: 
1)$C \leq F_n$ for all $n$; 
2) if $D$ is any element of  ${\cal L}$ such that 
$F_{n} \geq  D$ for all $n$, then $C \geq D$.}
   \item[$(\imath\imath)$] For $F_1, F_2\in {\cal L}$ and $F_1\leq F_2$ there
exists an element $P\in {\cal L}$ such that $P\leq F_1$ and
$P\vee F_1 =F_2$. This element is equal to $P=F_1^\bot F_2=
(\bbbone - F_1)F_2=F_2 - F_1$ and is unique with this property.
\end{enumerate}
An orthocomplemented lattice satisfying $(\imath),\
(\imath\imath)$ is called a {\it logic}.

\subsection{Symmetry and quantum mechanics}
Let the configuration manifold --- which we
shall always denote by $M$ --- be a $G$--space of a symmetry
group $G$. This means that an action of $G$ on $M$ is given,
i.e., to each element $g$ of $G$ there corresponds a
transformation $g$ of $M$ onto itself such that: 
\begin{itemize}
\item[(1)]{$e.u=u$,} \item[(2)]{$g_1 (g_2 .u)=(g_1 g_2).u$,
where $g_1 ,g_2\in G$, $u\in M$.} 
\end{itemize}

Some important assertions, in particular Mackey's Imprimitivity
Theorem, can be stated provided the group $G$ is locally compact
and separable (i.e. with countable basis of the topology). In
the following we shall restrict our considerations to the case
when $G$ is a \em finite-dimensional connected Lie group. \em It
is just these groups that very often appear in physical
applications. Let the manifold $M$ be also connected and smooth,
and the mapping $(g,u)\mapsto g.u$ be infinitely differentiable
($C^\infty$).

Now we would like to associate, to each symmetry transformation
$g$ of the configuration space $M$, a symmetry transformation of
the quantum mechanical description. To be specific, we introduce
two notions.

{\bf Definition 2.1}. An automorphism of a logic
${\cal L}$ is a one-to-one mapping $\alpha\colon {\cal
L}\rightarrow {\cal L}$ which satisfies 
\begin{itemize}
\item[(i)]{$\alpha(\bbbone)=\bbbone$,}
\item[(ii)]{$\alpha(F^\bot)=\alpha(F)^\bot$,}
\item[(iii)]{$\alpha(\bigvee_{n=1}^\infty
F_n)=\bigvee_{n=1}^\infty \alpha(F_n)$.} 
\end{itemize} 

{\bf Definition 2.2}. A convex automorphism of the set of states
${\cal W}$ is a one-to-one mapping $\beta\colon {\cal
W}\rightarrow{\cal W}$ with the following property: given
positive real numbers $c_1 , c_2 ,\dots$  such that
$\sum_{n}c_n =1$, then $\beta(\sum_{n}c_{n}U_n) =
\sum_{n}c_{n}\beta(U_n)$.

Now we can state the conditions on {\it symmetry transformations
of the quantum mechanical description} in the following form:
\begin{itemize} 
\item[(a)] There exists a homomorphism
$\alpha\colon g\mapsto \alpha (g)$ from the group $G$ into the
group of automorphisms of the logic ${\cal L}$, $\alpha
(g)\colon {\cal L}\rightarrow {\cal L}\colon F\mapsto F^g$.
\item[(b)] There exists a homomorphism $\beta\colon g\mapsto
\beta (g)$ from the group $G$ into the group of convex
automorphisms of the set of states ${\cal W}$, $\beta (g) \colon
{\cal W} \rightarrow {\cal W}\colon U\mapsto U^g$. \item[(c)]
The probability does not change under the symmetry
transformations, i.e. $${\rm Tr}(U^{g}F^{g})={\rm Tr}(UF),\qquad
U\in {\cal W}, \; F\in {\cal L}.$$ \item[(d)] Given a
projection--valued measure $F$ on a set $X$ (endowed with a
$\sigma$-algebra of measurable subsets) which corresponds to
measurements on the system with values in $X$, there exists a
homomorphism $\gamma\colon g\mapsto \gamma (g)$ from the group
$G$ into the group of measurable and one-to-one mappings from
$X$ onto itself. Every $\gamma (g)$ induces an automorphism of
the $\sigma$-algebra, ${\cal B}(X) \rightarrow {\cal B}(X)
\colon S \mapsto S^g$. We demand $$F(S^g)=F(S)^g.\eqno(2.1)$$
\end{itemize}

Automorphisms of the logic and convex automorphisms 
of the set of states are described by  {\it Wigner's theorem}:

{\bf Theorem 2.1}
(\cite{29}, Chap. VII.3). Let ${\cal H}$ be a separable
infinite--dimensional Hilbert space. Then: 
\begin{itemize}
\item[1)] All automorphisms of the logic ${\cal L}$ are 
   of the form 
   $$\alpha (F)=TFT^{-1}, \qquad F\in {\cal L},$$ 
   where $T$ is a fixed unitary or antiunitary operator in
   ${\cal H}$. Two such operators induce the same
   automorphism of the logic if and only
   if they differ by a phase factor. 
\item[2)] All convex automorphisms of the set of states
   ${\cal W}$ are of the form
   $$\beta (U)=TUT^{-1}, \qquad U\in {\cal W},$$ 
   where $T$ is a fixed unitary or antiunitary operator in
   ${\cal H}$. Two such operators induce the same convex
   automorphism of the set of states if and only if they
   differ by a phase factor.
\end{itemize}

Theorem 2.1 and conditions (a), (b) imply that to each
 action $g\in G$ a pair of operators $T(g)$, $T'(g)$ is
 associated, both being {\it unitary or antiunitary.} 
To fulfil condition (c), operators
$T(g)$, $T'(g)$ may differ by a phase factor at most. Hence
these operators can be identified, 
$$T(g)=T'(g), \qquad g\in G.$$ 

Since we consider only connected Lie groups, all operators
$T(g)$ will be unitary\footnote{This follows, on the one hand,
from the fact that the composition of two antiunitary operators
is unitary, and, on the other hand, from the fact that in some
neighbourhood $N$ of the unit element $e$ there exists 
$a\in N$ to each $b\in N$ such that $b=a^2$; it is well
known that $G$ is generated by the elements of $N$.}. 
Let us denote by $U({\cal H})$ the group of unitary
operators in ${\cal H}$ with
strong topology. The centre $Z$ of this group consists of
operators $z.\bbbone$, $z\in T^1$, where $T^1$ denotes 
the compact Lie group of complex numbers of unit modulus.
The quotient group $$P({\cal H})= U({\cal H})/Z$$ is called
the {\it projective group of the Hilbert space} ${\cal H}$.
The conditions (a) --- (c) can be summarized in one
requirement
\begin{quote} 
{\bf (abc)} There exists a homomorphism
$h\colon G\rightarrow P({\cal H})$.
\end{quote}
Moreover, we shall demand $h$ to be measurable. 
In this case $h$ is even continuous
(\cite{30}, Chap. VIII.5).

The condition (abc) can be reformulated with the use of the
notion of a projective representation. Let $\pi \colon U({\cal
H})\rightarrow P({\cal H})$ be the canonical homomorphism. For
each given homomorphism $h\colon G\rightarrow P({\cal H})$
 there exists a measurable mapping 
$V\colon G\rightarrow U({\cal H}),
\; V(e)=\bbbone, \; h=\pi\circ V$. 
The mapping $V$ is called the
{\it projective representation} of $G$. Two projective
representations $V$, $V'$ are called {\it equivalent}
if there exists
a measurable mapping $z \colon G\rightarrow T^1$ such that
$V'(g)=z(g)V(g)$. The homomorphism $h$ obviously determines the
projective representation uniquely up to this equivalence.

Given a projective representation $V$, there exists a measurable
mapping $$m \colon \ G\times G \rightarrow T^1$$ such that
$$V(a)V(b)=m(a,b)V(ab), \qquad a,b\in G.$$ The factor $m(a,b)$
is called a {\it multiplier} of $G$; by definition it fulfils
$$m(ab,c)m(a,b)=m(a,bc)m(b,c), \qquad m(a,e)=m(e,a)=1.$$ Two
multipliers $m$, $m'$ are {\it equivalent} if there exists a
measurable mapping $z\colon G\rightarrow T^1$ such that
$$m'(a,b)= z(ab)^{-1}z(a)z(b)m(a,b).$$ By definition a
multiplier is {\it exact} (or {\it trivial}), if it is
equivalent to 1. The set of all multipliers with pointwise
multiplication forms an Abelian group; trivial multipliers
form its invariant subgroup. The corresponding quotient group
is referred to as the {\it multiplier group} for $G$; we
shall denote it by ${\cal M}(G)$. 
\footnote{For details see \cite{30}, Chap. X.}

\subsection{Localization, systems of imprimitivity} 
The discussion of condition (d) of Sect. 2.2 was postponed
to this section, since its analysis requires the description
of a concrete measurement on the system. For localizable
systems the {\it position measurements} play a distinguished
r\^{o}le. Results of position measurements are points of the
configuration space $M$, i.e. $X=M$. So it is natural to
consider the {\it $\sigma$-algebra ${\cal B}(M)$ of Borel subsets of $M$} as the $\sigma$-algebra of measurable
sets~\footnote{$\sigma$-algebra
${\cal B}(M)$ is generated by open subsets of manifold $M$.}.

The starting point of Mackey's quantization and of
quantum Borel kinematics is the notion of localization
of a quantum system on a configuration manifold $M$.
It is mathematically modeled by a
{\it projection--valued measure} $E\colon S\mapsto E(S)$
mapping Borel subsets $S$ of $M$ ($S\in {\cal B}(M)$) into
projection operators $E(S)$ on a separable Hilbert space
${\cal H}$ subject to the usual axioms of localization. 
For convenience, these axioms are given below:
 \begin{eqnarray*} 
E(S_1 \cap S_2) & = & E(S_1) \cdot E(S_2),\\
E(S_1 \cup S_2) & = & E(S_1)+E(S_2)-E(S_1 \cap S_2),\\
E(\bigcup_{i=1}^{\infty} S_i) & = & \sum_{i=1}^{\infty}{E(S_i)} \quad
\mbox{for mutually disjoint} \quad S_i\in{\cal B}(M),\\
E(M) & = & \bbbone .
\end{eqnarray*} 

For a given subset $ S\in {\cal B}(M) $, the
projection $E(S)$
corresponds to a measurement which determines whether the system
is localized in $S$; its eigenvalues 1 (0) correspond to
situations when the system is found completely inside (outside)
$S$, respectively.

According to (d), each action $g\in G$ induces a Borel
transformation of $M$ onto itself. As already mentioned, the
other three conditions {\bf (abc)} imply the existence of a projective
representation $V$ of $G$. Hence (2.1) can be written in the
form 
$$E(g.S)=V(g)E(S)V(g)^{-1}, \qquad 
(S^g \equiv g.S).\eqno(2.2)$$

{\bf Definition 2.3.} 
A pair $(V,E)$ where $V$ is a
(projective) representation of a group $G$ and $E$ 
is a projection--valued measure on a $G$-space $M$, 
is called a {\it (projective) system of imprimitivity} 
for the group $G$, if
(2.2) holds for all $g\in G$, $S\in {\cal B}(M)$. 
Two projective systems of imprimitivity are {\it equivalent}
if the corresponding projective representations are
equivalent and if the projection--valued measures are equal.

\subsection{Quantization on homogeneous spaces} 
Stronger results
can be obtained if the symmetry group $G$ of $M$ is sufficiently
rich. More precisely, we shall turn our attention to homogeneous
spaces. By definition, $M$ is a {\it homogeneous G-space} if $G$
acts transitively on $M$, i.e., to each pair of points $u, u'\in
M$ there exists a transformation $g\in G$ such that $g.u=u'$.
Let us fix a point $u_0 \in M$. The isotropy subgroup of $u_0$
in $G$ will be denoted by $H$. It is well known that $H$ is a
closed Lie subgroup of the Lie group $G$. The space $G/H$ of
left cosets $gH$, $g\in G$, endowed with factor topology, can be
given a differentiable ($C^\infty$) structure, thus becoming a
smooth manifold, and the mapping $\pi \colon g\mapsto g.u_0$
induces a diffeomorphism of $G/H$ onto $M$ (\cite{21}, Chap.
II.3, II.4). Having identified $G/H$ with $M$, the group $G$
acts on $M$ in the natural way, $a\colon gH\mapsto agH$. The
quadruple $(G,\pi,M;H)$ can be viewed as a principal fibre
bundle. Let us note here that the requirement on $G$ to be
connected is not very restrictive provided $M$ is connected:

{\bf Proposition 2.2} (\cite{21}, Chap. II.4). Let $G$ be a
finite dimensional Lie group acting transitively on a connected
smooth manifold $M$ and let $G^0$ be the connected component of
unity in $G$. Then $G^0$ acts transitively on $M$, too.

As shown by Mackey \cite{11,26}, the {\it transitive systems of
imprimitivity} (i.e. based on homogeneous spaces $G/H$) can be
completely classified. The notions of irreducibility, unitary
equivalence, direct sum decomposition, etc., can be taken over
for the systems of imprimitivity in exact analogy with these
notions for (projective) unitary representations (\cite{26},
Chap. 1.2). In order to investigate questions of irreducibility,
direct sum decomposition, etc., a \em commuting ring \em ${\cal
C}(V,E)$ is considered, which consists of all bounded operators
in ${\cal H}$ commuting with $E(S)$, $V(g)$ for all $S\in {\cal
B}(M)$, $g\in G$. We have for instance the property that a
system of imprimitivity $(V,E)$ is irreducible if and only if
the ring ${\cal C}(V,E)$ consists of multiples of the unit operator
$\bbbone$ only (Schur's Lemma).

Now, following G.W. Mackey, we are going to describe 
the {\it canonical construction of transitive systems of
imprimitivity.}
Let $G$ be a locally compact group satisfying the second
axiom of countability, $H$ its closed subgroup. 
On the coset space $G/H$ there exists a quasi--invariant
measure defined on the $\sigma$-algebra of Borel subsets. 
 
A measure $\mu$ on $G/H$ is called {\it quasi--invariant}
with respect to the action of $G$, if for all $g\in G$ 
the measures $\mu$ and $\mu\circ g \colon S
\mapsto \mu (g.S)$ are mutually absolutely continuous. 
Moreover, all quasi--invariant $\sigma$-finite measures on
$G/H$ are mutually absolutely continuous (\cite{30}, Chap. VIII.4). 

We fix a measure $\mu$ from this class. 
Further, let $m$ be a multiplier of $G$ and let $L$ 
be a projective unitary representation of $H$ with
multiplier $m$  restricted to $H$ 
in a separable Hilbert space ${\cal H}^L$. 
Then we construct the Hilbert space ${\cal H}$ as
the space of vector--valued functions $\psi\colon G
\rightarrow{\cal H}^L$ satisfying 
\begin{itemize} 
\item[a)]
$a\mapsto <\psi (a),f>$ is a Borel function on $G$ 
for all $f\in {\cal H}^L$; 
\item[b)] $\psi(ah)=m(a,h)L_{h}^{-1}\psi(a)$, 
$h\in H$; 
\item[c)] $\parallel \psi \parallel < \infty$, where
$\parallel \cdot \parallel$ is the norm induced 
by the inner product 
$$(\psi,\psi')=\int_{G/H} <\psi(a),\psi'(a)>\, {\rm d}
\mu (u);$$ the integral is well-defined since, 
because of b), the inner product $<\psi(a),\psi'(a)>$ 
in ${\cal H}^L$ remains
constant on the left cosets $u=aH$. 
\end{itemize} 
Henceforth we shall identify two functions on $G/H$ which
are equal $\mu$-almost everywhere.

Then the projection--valued measure $S\mapsto E^{L}(S)$ on $G/H$
is canonically defined by
$$
[E^{L}(S)\psi](a)=\tilde{\chi}_{S}(a)\psi(a),$$ 
where
$$
\tilde{\chi}_{S}(a)=\left\{\begin{array}{lcr} 1 & \rm for \it
& aH\in S\\ 0 & \rm for \it & aH\notin S\\ \end{array}\right.$$
The projective $m$-representation of $G$, $g\mapsto
V^{L}(g)$, is given by 
$$
[V^{L}(g)\psi](a)=
\sqrt{\frac{{\rm d}\mu}{{\rm d}(\mu\circ g)}(g^{-1}u)} \, 
m(a^{-1},g)\, \psi(g^{-1}a),$$ 
where ${\rm d}\mu/{\rm d}(\mu\circ g)$ is the
Radon--Nikod\'ym derivative. The pair $(V^L,E^L)$ is called a 
{\it canonical system of imprimitivity} and its equivalence class
does not depend on the choice of a quasi-invariant measure
$\mu$.

{\bf Theorem 2.3} (The Imprimitivity Theorem \cite{11}). 
Let $G$ be a locally compact group satisfying the
second axiom of countability, $H$ its closed subgroup 
and $m$ a multiplier of $G$. 
Let a pair $(V,E)$ be a projective system of
imprimitivity for $G$ based on $G/H$ with multiplier~$m$.
Then there exists an $m$-representation $L$ of $H$ 
such that $(V,E)$ is equivalent to the canonical system 
of imprimitivity $(V^L,E^L)$. 
For any two $m$-representations $L$, $L'$ of the
subgroup $H$ the corresponding canonical systems of
imprimitivity are equivalent if and only if $L$, $L'$ are
equivalent. The commuting rings ${\cal C}(V^L,E^L)$ and
${\cal C}(L)$ are isomorphic.

The Imprimitivity Theorem shows how to obtain all systems of
imprimitivity up to unitary equivalence, provided the multiplier
group ${\cal M}(G)$ is known. More facts about the multiplier
group can be given in the case when $G$ is a connected and
simply connected Lie group. Then every multiplier is equivalent
to a multiplier of class $C^\infty$. These multipliers can be
expressed in the form $\exp (ip)$ where $p$ is called an {\it
infinitesimal multiplier}. Let us introduce a coboundary
operator $\delta$ on real skew--symmetric multilinear forms on the
Lie algebra ${\cal G}$ via $$\delta
p(X_1,\dots,X_n)=\sum_{k<j}(-1)^{k+j-1}
p([X_k,X_j],X_1,\dots,\hat{X_k},\dots,\hat{X_j},\dots,X_n),$$
where $p$ is any $(n-1)$-form. Then the Abelian group of
infinitesimal multipliers of $G$ is isomorphic to the second
cohomology group $H^{2}({\cal G},\bbbr)$.\\[3ex]
{\bf Examples}\\
\begin{itemize} 
\item[(i)] $H^{2}(\bbbr^s,\bbbr)$ is isomorphic to the
additive group of real skew--symmetric 2--forms on
 $\bbbr^s$; given a
2-form $p$, then $m \colon (x,y)\mapsto \exp [ip(x,y)]$ is a
multiplier; ${\cal M}(\bbbr^1)=\{1\}$. 
\item[(ii)] ${\cal
M}(T^s)=\{1\}$, $T^s = \bbbr^{s}/\bbbz^s$. 
\item[(iii)] If $G$ is a
connected and simply connected semi--simple Lie group, then
${\cal M}(G) = \{1\}$.
\end{itemize}

We note that any connected Lie group $G$ can be replaced by its
(connected and simply connected) universal covering Lie group
$\tilde G$. The action of $\tilde{g}\in \tilde G$ on $M$ is
given by $\tilde{g} \colon u \mapsto \pi (\tilde g).u$ where
$\pi \colon \tilde{G} \mapsto G$ is the covering homomorphism.
As Example 2.3 (see Sect.2.5) will show, the transition to the
covering group can lead to richer results with reasonable
physical interpretation.

{\bf Remark.} Detailed
descriptions of general foundations of quantum mechanics can be
found in \cite{29}, Chap. VI and VII; of systems of
imprimitivity in \cite{30}, Chap. IX, \cite{11} and \cite{7}; of
multipliers in \cite{30}, Chap. X.

\subsection{Infinitesimal action on a $G$-space} 
Let $G$ be a
connected Lie group and $M$ a (not necessarily homogeneous)
$G$--space. A one--parameter subgroup of $G$ is a 
one--dimensional
Lie subgroup including its parametrization, $a \colon \bbbr
\rightarrow $G$ \colon t \mapsto a(t)$. There is a 
one--to--one
correspondence between elements $A$ of a Lie algebra ${\cal G}$
and one--parameter subgroups $\{a(t)\}$ which can be expressed by
$a'(0)=A$ (\cite{23}, Chap. I.6.4). This correspondence can be
used to define the mapping $\exp \colon {\cal G} \rightarrow G
\colon A \mapsto a(1)$; then one has $a(t)=\exp (tA)$, and
$\exp$ is a local diffeomorphism at the unit element of $G$. To
each $A\in {\cal G}$ there corresponds a one--parameter subgroup
$\{a(t)=\exp (tA)\}$ and a flow on $M$, $(t,u)\mapsto a(t).u$;
the corresponding vector field on $M$ will be denoted by $D_A$.
If $F^u \colon G \rightarrow M$ is the mapping $g \mapsto g.u$
depending on $u\in M$, then obviously $D_{A}(u)=({\rm
d}F^{u})_{e}.A$. In the terminology of \cite {18}, an {\it
infinitesimal action} is the mapping ${\cal G} \rightarrow {\cal
X}(M) \colon A \mapsto D_A$, where ${\cal X}(M)$ denotes the
infinite--dimensional Lie algebra of smooth vector fields on $M$.

Let $N$ be the subgroup of ineffectively acting elements from
$G$. If $N=\{e\}$, then $G$ is said to act {\it effectively} on
$M$. $N$ is closed and normal, the factor group $G/N$ is a Lie
group acting effectively on $M$. Manifold $M$ can be considered
as a $G/N$-space if the action is given by $G/N \ni gN \colon u
\mapsto g.u$. In this way the ineffectively acting elements can
be eliminated\footnote{However, for $N$ discrete it does not
seem reasonable to eliminate $N$ in this way; see
Ex. 2.3.}.

{\bf Theorem 2.4} (\cite {18}, Chap. III.3.7).
The infinitesimal action $A\mapsto D_A$ is linear. For all
$A,B\in {\cal G}$ one has $$[D_A,D_B]=-D_{[A,B]}.$$ Hence the
image of this mapping is a finite--dimensional 
Lie subalgebra in ${\cal X}(M)$. 
The kernel is ${\cal N}$, the Lie algebra of the
group $N$ of ineffective elements from $G$.\\[1ex] 
The {\bf proof} of the first part is based on a straightforward
calculation (\cite{18}, Chap. III.3.7). For the last assertion we
observe that, if $D_A=0$ and $\phi^A$ is the corresponding flow,
then $a(t).u=\phi^{A}(t,u)=u$ so $\{a(t)\}\subset N$.

Now let us consider the opposite situation. Suppose we are given
a finite--dimensional Lie subalgebra ${\cal G}$ in ${\cal X}(M)$
such that all vector fields from ${\cal G}$ are complete. Let
$\tilde{\cal G}$ be the Lie algebra with the same vector space
as ${\cal G}$ but with a Lie bracket $[.,.]^{\sim}=-[.,.]$. If
$\tilde{G}$ is the connected and simply connected Lie group with
Lie algebra $\tilde{\cal G}$ ($\tilde{G}$ is unique up to
isomorphism), then according to \cite{18}, Chap. III.4.7, Theorem
6, to each $u\in M$ there exists an open neighbourhood $B_u$ and
a uniquely defined local action of $\tilde{G}$ on $B_u$ (i.e.,
an action defined only for elements from some neighbourhood
$U_e$ of the unity) such that the associated infinitesimal
action is identical with the mapping $\tilde{\cal G}\ni X\mapsto
X\mid_{B_u}$. The neighbourhood $U_e$ can be chosen
small enough for $\exp$ to be a diffeomorphism on it.
For $X\in \tilde {\cal G}$, $\phi^X$ ---
the corresponding flow, $v \in B_u$, $t\in \bbbr$
sufficiently small, we have $\exp (t.X).v=\phi^{X}(t,v)$. Since all vector fields $X\in \tilde{\cal G}$ are complete,
 $U_e$ can be chosen independently of $u\in M$. In this way
 we obtain a local homomorphism from $\tilde{G}$ into the
 group of diffeomorphisms of $M$. Since $\tilde{G}$ is
 connected and simply connected, the domain of the local
 homomorphism can be unambiguously extended to the whole
 $\tilde{G}$ (\cite{19}, Chap. II and VII).

\subsection{Examples} 
{\bf Example 2.1.} $M=\bbbr^1$, $G=\bbbr^1$ -- the
group of translations. Both the multiplier group ${\cal M}(\bbbr^1)$
and the isotropy subgroup are trivial. So in this case exactly
one irreducible system of imprimitivity exists (up to unitary
equivalence). Let ${\cal H}$ be a Hilbert space and $(Q,P)$ be a
pair of self--adjoint operators in ${\cal H}$ satisfying the
commutation relation $QP-PQ=i\hbar\bbbone$. Then if $E \colon S
\mapsto E(S)$ is the spectral projection--valued measure of $Q$,
and $V(t)=\exp (-itP/\hbar)$, then the commutation relation is
equivalent to the identity 
$$V(t)E(S)V(-t)=E(t+S).$$ 
So the Imprimitivity Theorem implies the Stone--von Neumann theorem (cf.
\cite{26}, Chap. 2.5).\\[1ex] 
{\bf Example 2.2.} $M=\bbbr^2$, $G=\bbbr^2$
-- the group of translations. To each element $A=(A_1,A_2)\in
{\cal G}=\bbbr^2$ there corresponds a vector field
$D_{A}=(A_{1}\partial/\partial{x_1})+
(A_{2}\partial/\partial{x_2})\in {\cal X}(\bbbr^2)$. The isotropy
subgroup is trivial. ${\cal M}(\bbbr^2)$ is isomorphic to the group
$\bbbr$ under addition: if $B\in \bbbr$, then 
$$m_B \colon \bbbr^{2}\times
  \bbbr^{2} \rightarrow U(1) \colon (x,y) \mapsto
  \exp[-i\frac{eB}{2\hbar}(x_{1}y_{2}-x_{2}y_{1})]$$  
is a multiplier ($e$ is an arbitrary fixed non-zero constant). The
inequivalent irreducible systems of imprimitivity $(V^B,E^B)$ are labelled by $B\in \bbbr$: the Hilbert space
 is ${\cal H}^B=
L^{2}(\bbbr^2,{\rm d}x_{1}{\rm d}x_{2})$ and we find $E^{B}(S) \colon \psi\mapsto \chi_{S}.\psi$ and 
$$
V^{B}(\exp (tA))\psi=\exp (-\frac{i}{\hbar}
 tP^{B}(D_A))\psi, \qquad t\in \bbbr,
$$ 
where $P^{B}(X)$ is a self--adjoint operator, 
$$
P^{B}(X)\psi =(-i\hbar X - e\alpha (X))\psi,
$$ 
$\alpha =(B/2)(x_{1}{\rm d}x_{2}-x_{2}{\rm d}x_{1})$, 
i.e. ${\rm d}\alpha =B{\rm d}x_{1}\wedge {\rm d}x_{2}$, 
and $\psi\in C_{0}^{\infty}(\bbbr^2)$.
The real number $B$ can be given physical meaning: 
a particle with electric charge $e$ moves on the plane
$\bbbr^2$ in an external magnetic field which is
perpendicular to the plane and has
constant value $B$ (the sign reflects the orientation).\\[1ex]
{\bf Example 2.3.} 
$M=S^1$, $G=$ U(1) --- the group of rotations of
the circle $S^1$. Both ${\cal M}(\mbox{U}(1))$ and the isotropy
 subgroup are trivial; hence there exists exactly one
 irreducible system of imprimitivity (up to unitary
 equivalence). Now let us replace $G=$ U(1) by its universal
 covering $\tilde{G}=\bbbr$. Then ${\cal M}(\bbbr)=\{1\}$
 and the isotropy subgroup $H=2\pi \bbbz$. The
 irreducible unitary representations of $H$ are labelled by
 elements $z\in \bbbr /2\pi \bbbz$: for $\Phi\in \bbbr$ 
let
$$
L^{\Phi} \colon 2\pi k\mapsto \exp (\frac{i}{\hbar} ke\Phi),
 \quad  k\in \bbbz ;$$ 
if $(e/\hbar)(\Phi -\Phi')\in 2\pi \bbbz$, then $L^{\Phi}=L^{\Phi'}$.

We shall describe the system of imprimitivity for given $\Phi\in \bbbr$. The Hilbert space ${\cal H}^{\Phi}$
 consists of (equivalence
classes of) functions $\psi \colon \bbbr \rightarrow \bbbc$ such that 
$$
\psi (x+2\pi k)=\exp(-\frac{i}{\hbar} ke\Phi)\psi(x),
\quad (k\in \bbbz)
$$ 
almost everywhere; the inner product is defined by
$$(\psi,\psi')=\int_{a}^{a+2\pi}\psi(x)\overline{\psi'(x)} {\rm d}x,
\qquad a\in \bbbr.$$ 
We have 
$$V^{\Phi}(t)=\exp(-\frac{i}{\hbar}tP^{\Phi}),$$
where $t\in \bbbr$ and 
$$P^{\Phi}=-i\hbar {\rm d}/{\rm d}x$$
 is self--adjoint. The mapping 
$$W \colon {\cal H}^{\Phi}\rightarrow {\cal H}^0 
\colon \psi(x)\mapsto
\exp(i\frac{e\Phi x}{2\pi \hbar}).\psi(x)$$ 
is unitary, ${\cal H}^0$ can be identified with 
$L^{2}(S^1, {\rm d}\varphi)$. We find 
$$
WP^{\Phi}W^{-1}=-i\hbar\frac{{\rm d}}{{\rm d} \varphi}
-\frac{e\Phi}{2\pi}, \qquad \varphi\in [0,2\pi).$$ 
A possible physical interpretation is connected with the
{\it Aharonov--Bohm effect} \cite{1}: 
a particle in $\bbbr^3$ with electric
charge $e$ is moving on the circle $x_{1}^{2}+x_{2}^{2}=1$,
$x_{3}=0$, and external magnetic flux $\Phi$ is concentrated
along the $x_{3}$-axis passing through the centre of the circle.
If $(e/2\pi\hbar)(\Phi -\Phi')$ is an integer, then the two
quantum kinematics with fluxes $\Phi\neq\Phi'$ lead to the same
observable results (e.g., the same interference pattern).\\[1ex]
{\bf Example 2.4.} 
$M=S^2$, the symmetry group $G={\rm SO}(3)$ is
replaced by the quantum mechanical symmetry group 
${\rm SU}(2)$
\cite{7} acting on $S^2$ in the usual way
$$
\left(
\begin{array}{lr} x_{3} & x_{1}-ix_{2}\\ x_{1}+ix_{2} & -x_{3}
\end{array} \right) \mapsto 
T \left( \begin{array}{lr} x_{3} &
x_{1}-ix_{2}\\ x_{1}+ix_{2} & -x_{3} \end{array} \right) T^\ast
,$$ 
where $T \in {\rm SU}(2)$ and points $(x_{1},x_{2},x_{3})
\in S^{2} \subset \bbbr^3$ ($\sum_{i}x_{i}^{2}=1$) are identified
with the matrices $\sum_{i} x_{i}\sigma_{i}$; $\sigma_{1}$,
$\sigma_{2}$, $\sigma_{3}$ are the Pauli spin matrices. The
isotropy subgroup $H$ of the north pole $(0,0,1)$ consists of the diagonal
matrices $\left( \begin{array}{lr} \tau & 0\\ 0 & \bar{\tau}
\end{array} \right) $, $\tau \in T^{1} \simeq {\rm U}(1)$, hence
$H \equiv {\rm U}(1)$.  If $T$ is parametrized by $\alpha ,\beta
\in \bbbc$, $\bar{\alpha}\alpha + \bar{\beta}\beta = 1$, 
$$
T =T(\alpha,\beta) = 
\left( \begin{array}{lr} \alpha & -\bar{\beta}\\
\beta & \bar{\alpha} \end{array} \right),$$
then the projection $\pi \colon {\rm SU}(2)
\rightarrow S^{2} \simeq {\rm SU}(2)/{\rm U}(1)$ is given by
$$
\pi (T) = T \sigma _{3} T^{*} = (2~{\Re}(\bar{\alpha}\beta),
2~{\Im}(\bar{\alpha}\beta),\bar{\alpha}\alpha-
\bar{\beta}\beta). $$
The quadruple $({\rm SU}(2),\pi,S^2;{\rm U}(1))$ constitutes a
non-trivial principal
bundle known as the {\em Hopf fibration}. We shall explicitly write
local trivializations of this bundle on sets $U_{n} = S^{2}
\backslash \{s\}$, $U_{s} = S^{2} \backslash \{n\}$, where
$n=(0,0,1)$ and $s=(0,0,-1)$ are the north and the south pole,
respectively. A local trivialization is determined by a selected
smooth local section. We choose (in spherical coordinates
$\vartheta$, $\varphi$) $$\rho_{n} \colon
(\vartheta,\varphi)\mapsto T(\cos (\vartheta /2), e^{i\varphi}
\sin (\vartheta /2)), \quad O\leq\vartheta <\pi,$$ $$\rho_{s}
\colon (\vartheta,\varphi)\mapsto T(e^{-i\varphi} \cos
(\vartheta/2), \sin (\vartheta /2)), \quad O<\vartheta
\leq\pi.$$ Since ${\rm SU}(2)$ is simple, connected and simply
connected, its multiplier group is trivial. The irreducible
representations $L^n$ of the isotropy subgroup $H={\rm U}(1)$
are labeled by integers $n\in \bbbz$, $L^{n} \colon \tau \mapsto
\tau^n$. In order to write down explicit expressions for the
operators $P(X)$ of generalized momenta, it is convenient to
work in the complex line bundle associated (via $L^n$) with the
principal bundle. Then the Hilbert space ${\cal H}$ of the
canonical system of imprimitivity corresponding to $L^n$
consists of measurable sections $\psi$ in the complex line
bundle; each section $\psi$ can be identified with a pair of
functions $(\psi_n,\psi_s)$, where $$\psi_{n,s} \in
L^{2}(U_{n,s},\sin\vartheta {\rm d} \vartheta {\rm d}\varphi);
\quad \psi_{s}(u)=e^{-in\varphi} \psi_{n}(u)$$ for almost all
$u \in U_{n} \cap U_{s}$. We choose $i\sigma_1$, $i\sigma_2$,
$i\sigma_3$ as basis of the Lie algebra ${\rm su}(2)$. The
element $-(i/2)\sigma_3$ induces the vector field $J_3$ on
$S^2$, $J_{3}= x_{1}(\partial /\partial x_{2})-x_{2}(\partial
/\partial x_{1})$. Further, the relation
$$
\exp(-\frac{i}{\hbar}tP(J_3))=
V(\exp(-\frac{i}{2}t\sigma_{3})), \quad t \in \bbbr,
$$ 
defines a self--adjoint operator $P(J_3)$ in ${\cal H}$.
Operators $P(J_1)$, $P(J_2)$ corresponding to vector fields
$J_1$, $J_2$ can be obtained by cyclic permutations. If for
$\lambda \in \bbbr^3$, $J=\sum_{i} \lambda_{i} J_i$ is a vector
field and if $\psi = (\psi_{n},\psi_{s})$ is a smooth local section,
then the self--adjoint operator $P(J)$ is determined by a pair of
operators $P_{n}(J)$, $P_{s}(J)$; a straightforward
calculation yields $$P_{n,s}(J) \psi (u) = (-i\hbar J -
e\alpha_{n,s}(J)- \frac{n\hbar}{2}(\lambda .u))\psi (u),$$ where
$u \in U_{n,s}$, $\lambda .u = \sum_{i} \lambda_{i} u_i$;
1-forms $i\alpha_{n,s}$ are the localizations on sets $U_{n,s}$ of the connection 1-form $i\alpha$ in the
associated bundle. In spherical coordinates
$$
\alpha_{n,s} =  \frac{n\hbar}{2e}(\pm 1 -
 \cos \vartheta) {\rm d} \varphi,$$
$$i\alpha_{s} =
i\alpha_{n} + \frac{\hbar}{e}({\rm d} e^{-in
\varphi})e^{in\varphi}.$$ (The constant $e$ is again arbitrary,
non-zero, but fixed.) On the intersection $U_{n} \cap U_{s}$ we find
$$
\beta = {\rm d}\alpha_{n} = {\rm d}\alpha_{s} =
\frac{n\hbar}{2e}\sin \vartheta {\rm d}\vartheta 
\wedge {\rm d} \varphi.
$$ 
We put 
$$g = \int_{S^2}\beta, \qquad {\rm i.e.} \quad 
eg = 2\pi n \hbar.
$$ 

The situation may have the following physical
interpretation: a particle with charge $e$ is moving in 
the magnetic field of the {\em Dirac monopole} with 
magnetic charge $g$ placed at the origin $O$ in $\bbbr^3$,
so on the sphere $S^{2} \subset \bbbr^3$ there is the
external magnetic field 
$B = B(u) = (g/4\pi)u$, $u \in S^{2} \subset \bbbr^3$. 
The relation $eg = 2\pi n\hbar$, $n \in \bbbz$, 
coincides with the {\it Dirac quantization
condition} \cite{5,GO}. Operators $P(J_k)$ are the well-known
conserved total angular momentum operators for a charged
particle moving in the Dirac monopole field.\footnote{A 
more detailed discussion of this example was given in
\cite{15}. The first treatments of the magnetic monopole
using a connection in a fibre bundle appeared in 
\cite{GP, 16, 17}.}\\[1ex]
{\bf Example 2.5.} 
$M=\bbbr^1$, $G$ is the group of orientation preserving
affine transformations of $\bbbr^1$.  Having identified $M$
with $\bbbr \times \{1\} \subset \bbbr^2$, $G$ acts on $M$ according to $$
\left(
\begin{array}{lr} a & b\\ 0 & 1 \end{array} \right) \colon
\left( \begin{array}{l} x \\ 1 \end{array} \right) \mapsto
\left( \begin{array}{c} ax+b \\ 1 \end{array} \right), \quad
a>0, b \in \bbbr.
$$ 
The isotropy subgroup $H$ of the origin $O$
consists of all matrices with $b=0$. The irreducible unitary
representations of $H$ are of the form $L^{c} \colon a \mapsto
a^{ic}$, $c \in \bbbr$. The Lie algebra ${\cal G}$ consists of all matrices 
$$
A = \left( \begin{array}{cc} A_{1} & A_{2} \\ 0 & 0
\end{array} \right), \qquad A_{1}, A_{2} \in \bbbr.
$$ 
An element $A \in {\cal G}$ induces a vector field 
$D_{A} = (A_{1}x+A_{2}){\rm
d}/{\rm d}x \in {\cal X}(\bbbr^1)$. Let us investigate ${\cal M}(G)$: $G$ is connected and simply connected, and a general
real skew--symmetric 2-form $p$ on ${\cal G}$ is of the form $p
\colon (A,B) \mapsto K(A_{1} B_{2}-A_{2}B_{1})$ for some constant
$K \in \bbbr$; but it is easily verified that $p(A,B)=q([A,B])$
where $q \colon A \mapsto KA_2$ is a 1-form on ${\cal G}$, so
${\cal M}(G)$ is trivial (see Sect. 2.3). Thus the irreducible
systems of imprimitivity are labelled by $c \in \bbbr$. Explicitly,
${\cal H}^{c}=L^{2}(\bbbr, {\rm d}x)$, $E^{c}(S)$ acts via
multiplication by indicator function $\chi_{S}(x)$, and
$$
[V^{c}(a,b)\psi ](x)=a^{-1/2}e^{ic \ln a}\psi
                             (\frac{x-b}{a}).
$$ 
The self--adjoint generalized momentum operators
 $P^{c}(D_A)$, $A \in {\cal G}$, defined by
$$
V^{c}(\exp(tA))=\exp(-itP^{c} (D_A)/\hbar),
$$ 
are of the form 
$$
P^{c}(X) = -i\hbar (X+\frac{1}{2} \mbox{ div}X)
           - \hbar c \, \mbox{ div}X, \qquad X=D_A.
$$

\section{Quantum Borel kinematics: localization}

\subsection{Generalized system of imprimitivity} 

Generally, for a given smooth manifold $M$ there is, 
{\it a priori}, no geometric
symmetry group. As indicated in \cite{4,8,Do-To,12}, 
the investigation of vector fields on
$M$ is a meaningful starting point. We denote by ${\cal X}(M)$
the Lie algebra of smooth vector fields on $M$, by ${\cal
X}_{0}(M)$ its subalgebra of compactly supported vector fields,
by ${\cal X}_{c}(M)$  the family of all complete vector fields,
${\cal X}_{0}(M) \subset {\cal X}_{c}(M)$. The flow $\phi^X$ of a
complete vector field $X$ represents a one--parameter group of
diffeomorphisms $\{\phi^{X}_t\}_{t\in \bbbr}$ of $M$, also
 called a {\it dynamical system} on $M$. 
And, vice versa, every dynamical
system is a flow of some (uniquely determined) complete vector
field $$[X\psi](u) = [{{\rm d} \over {\rm d}t}
(\psi\circ\phi^X_t)(u)]_{t=0}.$$ The family of dynamical systems
on $M$ will be denoted by $D(M)$. The following theorem
summarizes some well-known facts from differential geometry
\cite{24}, \cite{27}.\\[1ex] {\bf Theorem 3.1.} Let $f:M
\rightarrow M'$ be a diffeomorphism. Then $f'\colon {\cal
X}(M)\rightarrow {\cal X}(M')$, where $(f'.X)_{f(u)} = {\rm
d}f_{u}(X_u)$, is a Lie algebra isomorphism; the restriction
$f'\colon {\cal X}_{0}(M)\rightarrow {\cal X}_{0}(M')$ is also a
Lie algebra isomorphism; $f'\colon {\cal X}_{c}(M)\rightarrow
{\cal X}_{c}(M')$ is a bijection. The mapping $$f^D\colon
D(M)\rightarrow D(M') \colon \{\phi_t\}_{t\in
\bbbr}\mapsto\{f\circ\phi_{t}\circ f^{-1}\}_{t\in \bbbr}$$ is bijective
and $f^{D}(\phi^X) = \phi^{f'.X}$.

For every $\phi^{X}\in D(M)$ the manifold $M$ becomes a
$G$-space for the group $G=\bbbr$.
Attempting to generalize Mackey's quantization (Sects. 2.3,
2.4)
we require that there exist: a Hilbert space ${\cal H}$, a
projection--valued measure $E$ on $M$ and unitary
representations $V^X$ in ${\cal H}$ of the flows $\phi^{X}\in
D(M)$ such that 
$$
V^{X}(t)\,E(S)\,V^{X}(-t) =E(\phi^{X}_{t}.S),\eqno(3.1)
$$ 
where the objects ${\cal H}$, $E$ do not depend on the
choice of $\phi^X\in D(M)$. Equation (3.1)
is just a generalization of (2.2). Geometric shifts of
Borel sets $S \in {\cal B}(M)$ by flows $\phi^{X}_{t}$
along complete vector fields $X$ are represented in 
${\cal H}$ by unitary operators $V(t)$ such that (3.1)
 holds. 

Generalized momentum
operators can then be introduced via Stone's Theorem as
(essentially self--adjoint) infinitesimal generators $P(X)$
 of the one-parameter groups of unitary operators ---
shifts in ${\cal H}$  of the localized quantum system,
$$
V^{X}(t)=\mbox{exp}(-\frac{i}{\hbar}P(X)t)\,,
                  \qquad t\in \bbbr\,.
$$

\subsection{Quantum Borel kinematics}
The quantization of `classical' Borel kinematics
$({\cal B}(M),{\cal X}_{c}(M))$ thus requires \cite{8}
the imprimitivity condition (3.1) for the unitary
representation of the flow of each complete vector field
individually. Then we can state

{\bf Definition 3.1.} {\em Quantum Borel kinematics} is a pair
$(V,E)$, where $E$ is a projection--valued measure on $M$
 in a separable Hilbert space ${\cal H}$, and $V$ associates
 with each $\phi^X\in D(M)$ a homomorphism 
$V^X\colon \bbbr\rightarrow U({\cal H})$ such that the
 following conditions are satisfied:
\begin{itemize} 
\item[1)]{Equation (3.1) holds for all $t\in \bbbr$,
$X\in{\cal X}_{c}(M)$, $S\in {\cal B}(M)$;} 
\item[2)]{The
mapping $P \colon X\mapsto P(X)$ from the Lie algebra ${\cal
X}_{0}(M)$ into the space of essentially self--adjoint operators
with common invariant dense domain in ${\cal H}$ is a Lie
algebra homomorphism: $$P(X+aY)=P(X)+aP(Y),\eqno(3.2)$$
$$[P(X),P(Y)]=-i\hbar\,P([X,Y];\eqno(3.3)$$} 
\item[3)]
{{\it Locality condition}. If two flows $\phi^{X_i} \in D(M)$,
$i=1,2$, after restriction on the set $(-a,a)\times S$, $a>0$,
$S \in{\cal B}(M)$, coincide, then the mappings $\bbbr\times{\cal
H}\to {\cal H}\colon (t,\psi)\mapsto V^{X_i}(t)\,\psi$ coincide on the
domain $(-a,a)\times {\cal H}_S$, where ${\cal H}_S$ is the
subspace of ${\cal H}$ projected out by $E(S)$.} 
\end{itemize}
If in 2) only linearity (3.2) is required, we shall call
 $(V,E)$ a {\it generalized system of imprimitivity}
for $D(M)$.\footnote{The family $D(M)$ acts transitively on
 $M$ (we suppose $M$ to be connected, without boundary):
for any two points $u,v \in M$ there exists
$X \in {\cal X}_{0}(M)$, $\phi^{X} \in D(M)$ such that
$\phi^{X}(1,u)=v$.} It will describe {\it quantum Borel
 kinematics with external gauge field.}

It follows from condition 3) that, if $V^X$ is known for all $X
\in {\cal X}_{0}(M)$, then it is determined for all $X\in
 {\cal X}_{c}(M)$; 3) further implies that $P(X)$ are
{\it  differential operators}. Condition 2) may sometimes be too restrictive since
(3.3) excludes a non-vanishing external gauge field on M; in
this connection see  \cite{Do-To,DTtorus,14} and also Ex. 2.2,
Ex. 2.4 and Sect. 4.7.

The projection--valued measure $E$ induces in a natural way
 a {\it quantization} $Q$ of classical (smooth) real
 functions $f\colon M\rightarrow \bbbr$ on configuration
 space (e.g. coordinate functions, potentials, etc.). Not
  necessarily bounded,
self--adjoint quantum position operators $Q(f)$ are uniquely
determined by their spectral decompositions 
$$
Q(f) =
\int_{-\infty}^\infty\,\lambda\,{\rm d}E_\lambda^f\,,
$$ 
where the spectral function $E_\lambda^f$ is given by the
 spectral measure
$E^{f}(\Delta) = E(f^{-1}(\Delta))$ on subsets $\Delta =
(-\infty,\lambda)$ of $\bbbr\,$. Equation (3.1) is then replaced by
$$V^{X}(t)\,Q(f)\,V^{X}(-t) = Q(f\circ\phi^{X}_{-t}),\eqno(3.4)$$
where $f\in C^{\infty}(M,\bbbr)$, and implies a generalization of
the Heisenberg commutation relations in terms of
coordinate-independent objects 
$$
[Q(f),P(X)]=i\hbar\,Q(X.f)\quad
\hbox{\rm on} \quad {\cal D}\subset{\cal H}.\eqno(3.5)
$$ 
It is assumed that operators $P(X),\ Q(f)$ have a common
invariant dense domain ${\cal D}$ in ${\cal H}$. If an
obvious relation
$$
[Q(f),Q(g)]=0\quad \hbox{\rm on}\quad{\cal D}\eqno(3.6)
$$ 
for all $f,g \in C^{\infty}(M,\bbbr)$
is still added, then (3.6), (3.5) and (3.3) define a 
{\it Schr\"{o}dinger system} in the sense of \cite{12}.

 We can say that {\it Borel quantization}
on a smooth configuration manifold $M$ associates 
the generalized position $Q(f)$ and momentum operators
$P(X)$ with smooth functions $f \in C^{\infty}(M,\bbbr)$ and
smooth vector fields $X \in {\cal X}(M)$, respectively.
These quantum kinematical observables on $M$ are globally 
defined, hence Borel quantization incorporates the global
structure of $M$. 

{\bf Remark.} The natural infinite--dimensional Lie
algebra structure (3.6), (3.5) and (3.3) of quantum Borel
kinematics should be compared with the {\it non--relativistic
local current algebra} for a Schr\"odinger second quantized
field over $M = \bbbr^3$ studied in \cite{31}:
\begin{eqnarray*}
\left[ \rho (f_1), \rho (f_2) \right] & = & 0, \\
\left[ \rho (f), J(X) \right]  & = & i \hbar \rho (X f), \\
\left[ J(X), J(Y) \right]  & = & -i \hbar J([X, Y]),
\end{eqnarray*} 
where $f, f_i \in C^{\infty}(\bbbr^3, \bbbr)$,
$X, Y \in {\cal X}(\bbbr^3)$. There is an apparent
algebraic correspondence of the local density operator
$\rho (f)$ with $Q(f)$ and the local current operator
$J(X)$ with $P(X)$. In both cases the representations 
of the algebraic structures yield possible quantum kinematics.
However, in contrast to quantum Borel kinematics,
where $Q(f)$ is a multiplication operator,
local current algebra is more general, since also
representations where $\rho (f_{1}f_2)$ is not equal to
$\rho (f_1)\rho (f_2)$ are admitted.

\subsection{Quasi-invariant measures, projection--valued
measures} 

The question of existence and uniqueness of a measure
which is {\it quasi-invariant} with respect to all
diffeomorphisms $\phi_1\colon u\mapsto\phi(1,u)$ for which
$\phi\in D(M)\,$, is answered by

{\bf Theorem 3.2.} The family of quasi-invariant measures on
${\cal B}(M)$ is non-empty and, moreover, all measures in this family are mutually
equivalent and form a unique invariant measure
class.\footnote{The invariant measure class is called the {\it Lebesgue measure class}.} 
After completion, those subsets in $M$ which have measure
zero are exactly measure zero sets in the sense of Lebesgue.
 \\[1ex]
{\bf Proof:} The fact that the family
of sets of zero measure in the sense of Lebesgue is invariant
under diffeomorphisms is well known \cite{20}. The existence
part of the theorem can be seen as follows. Having embedded $M$
in $\bbbr^m$ (Whitney's Theorem), we can consider a tubular
neighborhood $M^\varepsilon$ of $M$ in the normal bundle
(\cite{20}, Chap. 2.3). Denoting by $\pi\colon
M^\varepsilon\rightarrow M$ the associated submersion, we can
define $\mu(S)=\lambda^{m} (\pi^{-1}(S))$ for $S\in{\cal B}(M)$,
where $\lambda^m$ denotes the Lebesgue measure in $\bbbr^m$; then
$\mu$ is quasi-invariant. The assertion about uniqueness of the
invariant measure class for $M=\bbbr^n$ follows from the fact that
the family of diffeomorphisms $\phi_1$ includes all translations
and the assertion for the group of translations is known
(\cite{30}, Chap. II.3). In general, $M$ can be covered by a
countable family of open sets, each of which is diffeomorphic to
$\bbbr^n$ and so the assertion is true again. $\Box$

Measure $\mu$ will be called {\it differentiable} if the
mapping
$\bbbr \rightarrow \bbbr \colon t \mapsto \mu(\phi_{t}.S)$
is smooth for all $S \in {\cal B}(M)$, $\phi \in D(M)$. 
For instance, measure $\mu$ used in the proof of Theorem 3.2
is differentiable. Every manifold is locally orientable;
having fixed an orientation on an open set $U \subset M$
(dim $M = n$), then to every
differentiable measure $\mu$ exactly one $n$-form $\omega$
exists on $U$ such that
$$
\int_{U} f.\omega = \int_{U}f(u){\rm d}\mu(u)
\quad \hbox{\rm for all} \; f \in
C_{0}^{\infty}(U);
$$
$\omega_{u}(X_1,\dots,X_{n})>0$ for every
positively oriented basis in $T_{u}U$. 
For each $X \in {\cal X}(U)$ we define a function 
$\mbox{div}_{\mu}X$ on $U$ by 
$$
\mbox{div}_{\mu}X.\omega = k! \; 
{\rm d}({\rm i}_{X} \omega),     \eqno(3.5)
$$ 
where the $(n-1)$-form ${\rm i}_{X} \omega$ is
defined by 
$$({\rm i}_{X} \omega)(X_1,\dots,X_{n-1})=\omega
(X,X_1,\dots,X_{n-1}).
$$ 
Equivalently, if $\phi = \phi^{X} \in
D(M)$, then 
$$\frac{{\rm d}}{{\rm d}t}(\frac{{\rm d}\mu}
{{\rm d}(\mu \circ \phi_{t})})(\phi_{-t}.u)\mid _{t=0}= 
-\mbox{div}_{\mu}X.\eqno(3.6)
$$ 
In local coordinates, if 
$\omega = \rho(u){\rm d}u_{1} \wedge \dots \wedge 
{\rm d}u_{n}$, $\rho > 0$,
then 
$$\mbox{div}_{\mu}X = X.\ln \rho + \sum_j 
\frac{\partial X_j}{\partial u_j}.
$$

The structure of projection--valued measures on $M$ is
well known. According to \cite{30}, Chap. IX.4, we have
a canonical representation of a localized quantum system:
Let $E$ be a projection--valued measure on $M$ in a
separable Hilbert space ${\cal H}$. Then
there exist two sequences
$\{{\cal K}_r\}$, $\{\nu_r\}$,
$r=\infty,1,2,\dots$, the first one consisting of Hilbert
spaces, the other of measures on $M$ such that 
$\mbox{dim}\; {\cal K}_r=r$ and  $\nu_r$, $\nu_s$ are
mutually singular\footnote{i.e., there
exist $S_r,S_s\in{\cal B}(M)$, $S_r\cup S_s=M$, 
$S_r\cap S_s=\emptyset$, $\nu_{r}(S_s)=\nu_{s}(S_r)=0$.} 
for $r\neq s$.
The projection--valued measure $E$ is unitarily equivalent
 to the measure $E^0$ which acts via multiplication by
 indicator functions of subsets in the Hilbert space 
${\cal H}^0=\bigoplus_{r} {\cal H}_r$, where ${\cal H}_r$
are the Hilbert spaces of vector--valued functions from $M$
to ${\cal K}_r$, ${\cal H}_r=L^{2}(M,{\cal K}_{r},\nu_r)$.
The measures $\nu_r$ are determined uniquely up to
equivalence. If only one $\nu_r$ is
non-zero, the projection--valued measure $E$ is called 
{\it homogeneous}. Due to the transitivity of actions of the
 family $D(M)$ the following theorem holds (for details see
 \cite{30}, Chap. IX.5, IX.6):

{\bf Theorem 3.3.} If $(V,E)$ is a generalized system of
imprimitivity for $D(M)$, then $E$ is homogeneous. 
The unique non-zero measure $\nu_r$ belongs to the
Lebesgue measure class on $M$.

Thus the canonical representation of a localized quantum
system on $M$ involves a smooth measure, i.e. a measure
induced by the Lebesgue measure of the coordinate charts.
An $r$--homogeneous localized quantum system of degree $r > 1$
can be interpreted as a quantum system with {\it internal
degrees of freedom}; a 1--homogeneous localized quantum
system will be called {\it elementary} as its $E$'s are
related to elementary spectral measures. We shall need also

{\bf Theorem 3.4.} (\cite{30}, Chap. IX.2). 
Let $\mu$ be a Borel measure on $M$, ${\cal K}$ a Hilbert
 space, ${\cal H} = L^2 (M, {\cal K},\mu)$, $E$ a
 projection--valued measure on 
${\cal B}(M)$ acting in ${\cal H}$ via multiplication by
indicator functions. Then any bounded operator $B$  in
${\cal H}$ commuting with $E(S)$ for
all $S \in {\cal B}(M)$ ($B \in {\cal C}(E)$) is of the form
$$
B \colon \psi (u) \mapsto b(u).\psi (u),$$ 
where $b$ is a Borel mapping from $M$ into the
space of bounded operators in ${\cal K}$ such that 
$\mbox{sup}_{u \in M} \mid b(u) \mid < \infty$. 
Function $b$ is determined by $B$ uniquely on $M$ modulo a
set of $\mu$-measure zero.

\subsection{An example} 
Quantum Borel kinematics are rather
diverse (even for trivial configuration space $\bbbr^n$), as
the following example \cite{14} shows. 
Let $M=\bbbr^n$ with a fixed basis. To every vector field
$X=\sum_{k}^{} X_{k}\,\partial/\partial x_{k}$ we relate a
matrix--valued function 
$$
A(X)\colon \bbbr^{n}\rightarrow
\bbbr^{n,n},\;\; [A(X)_u]_{i,j}={\partial X_{i}(u)\over \partial x_j}.$$ 
It is straightforward to verify
$$
[A(X),A(Y)]-X.A(Y)+Y.A(X)=-A([X,Y]).\eqno(3.7)$$ 
Let $L$ be a skew-Hermitean representation of the Lie
algebra $\mbox{gl}^{+}(n,\bbbr)$ in a Hilbert space 
${\cal H}^L$. We define operators $Q(f)$, $P(X)$ in 
${\cal H}=L^{2} (\bbbr^n,{\cal H}^L,
{\rm d}x_1\dots{\rm d}x_n)$ by
$$
Q(f)\psi=f.\psi,\quad
P(X)\psi=-i\hbar[X+{1\over2}\mbox{div}X-L(A(X))].$$ 
Then using (3.7) and the identity $X.L(A(Y))=L(X.A(Y))$, 
the pair $(P,Q)$ can be shown to be a quantum Borel
kinematic. Choosing the representation $L$ in 
${\cal H}^L=\bbbc$ to be given by
$$
L(A)=-ic\,\mbox{tr}A,$$ 
where $c$ is a real constant, we obtain
$$
L(A(X))=-ic\,\mbox{div}X.$$ 
This is just an example of the divergence term which we
shall encounter in Sect. 4.7.

{\bf Remark.} Consider the surjective mapping 
$$
p\colon\mbox{gl}(n,\bbbr)\rightarrow 
\mbox{sl}(n,\bbbr)\colon A \mapsto A-{1\over n}
(\mbox{tr}A)\bbbone_n,$$
where $\bbbone_n$ is the unit $n\times n$-matrix. 
This mapping permits to associate with every representation
$L'$ of $\mbox{sl}(n,\bbbr)$ a representation $L=L'\circ p$
of $\mbox{gl}(n,\bbbr)$.
Then our mapping $P$ is the infinitesimal form of a
representation of the group of diffeomorphisms of $\bbbr^n$
induced from $\mbox{SL}(n,\bbbr)$ (see \cite{31} for $n=3$).
For general results concerning the ``divergence--like'' terms
in the framework of systems of imprimitivity for the group
of diffeomorphisms, see \cite{St88}.

\section{Quantum Borel kinematics: external gauge fields}

\subsection{External magnetic field} 
In order to motivate our construction of quantum Borel
kinematics with external field via generalized systems
of imprimitivity for $D(M)$,
let us consider quantum kinematics on $\bbbr^3$, 
for a charged particle in an external magnetic field 
$\vec{B}$. Since
$\mbox{div}\vec{B} = 0$, 
the Poincar\'{e} lemma implies that
there exists a vector potential $\vec{A}$ 
such that $\vec{B}= \mbox{rot}\vec{A}$. The Hilbert space is
 ${\cal H}= L^{2}(\bbbr^3,{\rm d}^{3}x)$, 
and the Hamiltonian 
$$
H = - \frac{\hbar^2}{2m}\sum_j
(\frac{\partial}{\partial x_j}-i\frac{e}{\hbar}A_j)^2.$$
If another potential $\vec{A'}$, $\mbox{rot}
\vec{A'}=\vec{B}$, is chosen, then according to the
Poincar\'{e} theorem there exists a
real function $\lambda$ such that $\vec{A'}=\vec{A} + 
\mbox{grad}\lambda.$ The Hamiltonian  $H$ is transformed as
$ H \mapsto  H'=W H W^{-1},$ where $W$ is the unitary
mapping 
$$
W : {\cal H \rightarrow H'}: \psi \mapsto
 \mbox{exp}(\frac{ie}{\hbar}\lambda )\psi.$$

This common quantum mechanical scheme can be reformulated in
geometric language \cite{16} : ${\cal H}$ is the space of
measurable sections in the trivial Hermitian complex line
bundle $\bbbr^3 \times \bbbc^1$ associated with the principal 
bundle $\bbbr^3 \times T^1$; 
$$i(e/\hbar )\alpha,  \qquad
\alpha = \sum{A_j \d x_j},$$
is a localized connection 1--form; 
$$ i(e/\hbar )\beta,  \qquad
\beta = \d\alpha = B_1 \d x_2 \wedge \d x_3 +
\mbox{cycl.}$$ 
is the curvature 2--form; 
$$
\vec x \mapsto \mbox{exp}(\frac{ie}{\hbar}
\lambda(\vec x))$$ 
is a transition function in the principal bundle
relating two different trivializations.

In general, if a magnetic field is given by a closed 2--form
$\beta$ on manifold $M$, a vector potential 1-form $\alpha$
 such that $\beta = \d\alpha$ need not exist. However,
 following \cite{16}, one can always define vector
 potentials $\alpha_k$ locally, i.e. on open sets $U_k$
 (diffeomorphic to $\bbbr^n$) such that \{$U_k$\} is an open
 covering of $M.$ We require that vector
potentials $\alpha_k,\alpha_j$ be related on the intersection
$U_k\cap U_j$ by a gauge transformation 
\[ 
i\alpha_j = i\alpha_k
+\frac{\hbar}{e}(\d~\mbox{exp} (\frac{ie}{\hbar}\lambda)) \mbox{exp} (-\frac{ie}{\hbar}\lambda)=
i(\alpha_k + \d \lambda),\]
where $\exp (\frac{ie}{\hbar}\lambda (\vec x))$ is a
transition function.
Then $\d \alpha_j =\d \alpha_k.$ In this way we shall
construct a principal bundle with typical fibre $T^1$ and
 connection \{$i\alpha_j$\}.

\subsection{Construction of a class of generalized systems
 of imprimitivity}
We choose a measure $\mu$ from the Lebesgue measure class on $M$
and four objects $(P, G, \Gamma,L),$ where: $P\equiv (P,\pi,M,G)$
(or shortly $P(M,G)$) is a principal bundle over $M$; its
typical fibre $G$ is an Abelian or compact Lie group; $\Gamma$ is
a connection in $P$, and $L$ is a unitary representation of $G$
in a finite-dimensional Hilbert space $\cal K$ with inner product
$<.,.>.$ We construct a separable Hilbert space $\cal H$ like in
Sect. 2.3, consisting of vector--valued functions $\psi : P
\mapsto \cal K,$ such that 
\begin{enumerate}
\item $x \mapsto <\psi (x),f>$ is a Borel function on $P$
 for all $f\in \cal K,$
\item $\psi (xg)=L^{-1}_g \psi (x), g \in G$ ($\psi$ is an
  equivariant function),
\item $\|\psi \| < \infty,$ where $\|.\|$ is the norm
 induced by the inner product
\[
(\psi, \psi')=\int_M{<\psi(x),\psi'(x)>\d \mu (u)}.\]
\end{enumerate}
The integral is well defined as the integrand remains
constant on the fibres. Two functions $\psi, \psi'$ are
identified if they coincide almost everywhere.The
projection--valued measure $E$ on $M$ is defined via
multiplication by indicator functions:
$$
E(S) \psi=\tilde \chi_S \psi, \quad 
\tilde \chi_S (x)=\left\{
\begin{array}{ll}
 1,& \pi (x)\in S\\0,& \pi (x) \notin S
\end{array} \right .\eqno(4.1) 
$$ 
For $\phi \in D(M)$  we define 
the unitary representation of the additive group $\bbbr$
$$
[V^{\phi}(t)\psi](x)=
\sqrt{\frac{\d\mu}{\d(\mu\circ\phi_t)}(\phi_{-t}u)} 
\psi (\tilde\phi_{-t}x), \quad u=\pi (x).\eqno(4.2)
$$
Here $\tilde\phi$ denotes the horizontal lift
of the flow $\phi$ on $M$.

It is easily verified that the constructed pair $(V,E)$ is a
generalized system of imprimitivity for $D(M)$ in the sense
of Definition 3.1. The linearity of mapping $P$ required by
this definition will be investigated in Sect. 4.3 
(see Eq. (4.7)).

The equivalence class of $(V,E)$ does not depend on measure
$\mu$; if $\mu'$ is another measure from the Lebesgue class,
then ${\cal H \rightarrow H'}: 
\psi\mapsto\sqrt{{\d\mu}/{\d\mu'}} \, \psi$ is
the desired unitary mapping. For simplicity we shall suppose
$\mu$ to be differentiable. We assumed $G$ to be Abelian or
compact in order to deal only with finite--dimensional
unitary representations $L$ of $G$. For dim $\cal{K} =r$ 
we identify $\cal K$ with $\bbbc^{r}$ endowed with the standard
inner product. We say that the pair $(V,E)$ is a 
\em generalized system of imprimitivity \em
specified by the quadruple $(P,G,\Gamma,L).$

We note that, for two diffeomorphic manifolds, there is a
one--to--one correspondence between the equivalence
classes of generalized systems of imprimitivity constructed
in this way.

\subsection{Construction in the associated vector
bundle}
The associated vector bundle $(F,\bar{\pi},M;\bbbc^r)$ will
be constructed in the standard way (see e.g. \cite{24,27}).
 We introduce equivalence relation
$$
(x;\xi)\sim (xg;L_g^{-1}\xi), \qquad g\in G,$$
on $P\times \bbbc^r$ and put $F=P\times \bbbc^r /\sim$; the
projection is $\bar\pi :F\rightarrow M:[x;\xi]\mapsto \pi(x).$
Each fibre becomes an $r$-dimensional Hilbert space:
$$
\lambda [x;\xi]+[x;\xi']=[x;\lambda\xi+\xi'],$$
$$
<[x;\xi],[x;\xi']>=<\xi,\xi'>, \qquad \pi(x)=u.$$
The inner product is well--defined since $L$ is unitary. For
 the same reason we can relate a unitary mapping $\hat x$ to
 each $x \in P:$
$$
\hat x:\bbbc^r \rightarrow F_{\pi(x)} : \xi \mapsto [x;\xi]$$
and we have $\widehat{xg}=\hat x\circ L_g.$

Let $\bar{\cal H}$ be the Hilbert space of measurable
 sections in the associated vector bundle, having finite
 norm induced by the inner product
$$
(\sigma, \tau)=\int_{M}<\sigma(u),\tau(u)> 
\d \mu(u).     \eqno(4.3) $$
We can define a unitary mapping $T:{\cal H}\rightarrow
 \bar{\cal H}: \psi \mapsto \sigma $ in a natural way:
$$
\sigma (u)=[x;\psi (x)], \qquad \pi(x)=u.$$
The definition of $T$ is correct since 
$(x;\psi(x))\sim (xg;L_g^{-1}\psi(x))$ and 
$L_g^{-1}\psi (x)=\psi (xg).$ 
The inverse mapping $T^{-1}:\sigma \mapsto \psi$ is given by
$\psi (x)=\hat x^{-1}\circ \sigma (\pi(x)).$ Having
 performed the unitary transform $T$ we replace $(V,E)$ by a
 generalized system of imprimitivity $(\bar V,\bar E)$ in
 $\bar{\cal H}$. We shall compute explicit expressions.

Clearly, $\bar E$ acts via multiplication by indicator
functions
$$
\bar E (S)\sigma =\chi_{S}. \sigma.\eqno(4.4)$$
In order to express $\bar V$ we must first describe induced
connection in the associated vector bundle $F$. Having
 Hermitian structure on the fibres, we consider only
{\it  Hermitian connections} on $F,$ i.e. connections for
 which all linear isomorphisms
$C(u_t):F_{u_0} \rightarrow F_{u_t}$ (shortly $C_t,$ see
 below), which belong to curves $u_t$ in $M$, are unitary.
 As is well known, there is a one-to-one correspondence
 between Hermitian connections and Hermitian covariant
 derivatives in $F.$ A covariant derivative $\nabla$ acting
on smooth sections, 
$ \nabla_X:\mbox{Sec} F \rightarrow \mbox{Sec} F$,
$X \in {\cal X} (M),$ is Hermitian, if it satisfies (in
addition to four conditions \cite{24} defining the covariant
derivative) the identity
$$
X<\sigma,\tau>=<\nabla_X \sigma,\tau>+<\sigma, \nabla_X
\tau>;$$ 
Sec $F$ denotes the linear space of smooth sections in
$F.$

A connection $\Gamma$ in $P$ induces a Hermitian connection
 in $F;$ given a piecewise smooth curve $u_t$ in $M$ and 
its lift $x_t$ in $P,$ then $C_t\quad: F_{u_0} \rightarrow
F_{u_t}$, $C_t=\hat{x_t}\circ \hat{x_0}^{-1}$ is the desired
unitary mapping. Its definition does not depend on the
starting point $x_0,$ since $(\widehat{x_t g})\circ 
(\widehat{x_0 g})^{-1}=\hat{x_t}\circ \hat{x_0}^{-1},$ 
and $x_t g$ is another horizontal lift of the curve $u_t.$
 Now the {\it Hermitian covariant derivative} is defined by
 the limit
$$
(\nabla_X \sigma )(u_0)=\lim_{t \to 0}\frac{1}{t}[C_t^{-1}
\circ \sigma(u_t)-\sigma(u_0)],\quad
 \dot{u_0}=X.\eqno(4.5)$$

We are now in the position to give explicit formula for
 $\bar V^{\phi}(t)$:
$$\bar V^{\phi}(t)\sigma(u)=
C(\phi_{-t}u)^{-1}[\sqrt{\frac{\d\mu} {\d
(\mu \circ\phi_t)}}\sigma]
(\phi_{-t}u).\eqno(4.6)$$
Let $\mbox{Sec}_0 \; F$ denote the linear space of smooth
 and compactly supported sections in $F;$ the subspace 
$\mbox{Sec}_0 \;F$ is dense in $\bar{\cal H}.$ As before, 
a self--adjoint operator $\bar P(X)$ is defined by 
$$
\bar V^{\phi^X}(t)=\mbox{exp}(-\frac{i}{\hbar}
\bar P (X)t);$$ 
using (4.6), (4.5) and (3.6) we find
$$
\bar P(X)\sigma= -i\hbar(\nabla_X + \frac{1}{2}
\mbox{div}_{\mu}X)\sigma , 
\quad \sigma \in \mbox{Sec}_0 \;F.\eqno(4.7)$$
Note that expression (4.7) implies that $\bar P \sim P$ is
linear in $X \in \cal{X}_{0}(M).$

Using unitary representation $L$ we can associate a
principal bundle $\tilde P(M,\mbox{U}(r))$ to 
the principal bundle $P(M,G).$ The construction is similar 
to that for $F$. 
On $P \times \mbox{U}(r)$ we introduce the equivalence relation
$(x;a)\sim (xg;L_g^{-1}a)$, $g \in G$, and put 
$\tilde P=P \times \mbox{U}(r)/\sim$,
$\tilde\pi:[x;a]\mapsto \pi (x) $. Lie group $\mbox{U(r)}$ 
acts on $\tilde P$ via $[x;a].b=[x;ab]$,
$b \in \mbox{U}(r)$. The mapping $ f: P \rightarrow \tilde P:x
\mapsto [x;e]$ is a bundle homomorphism --- we have 
$ f(xg)=f(x)L_g.$ With the help of
this homomorphism we can transform the connection 
$\Gamma$ in $P$ into a connection $\tilde\Gamma $
in $\tilde P,$ as described in the following theorem.

{\bf Theorem 4.1}. (\cite{27}, Chap. II.5; \cite{24}). 
Let $f:P(M,G) \rightarrow \tilde P(M,\tilde G)$ be a
principal bundle homomorphism and $\Gamma$ a connection in
$P.$ Then there exists a unique connection $\tilde\Gamma $
in $ \tilde{P}$ such that the tangent mapping $f_{\ast}$
maps every horizontal subspace of connection $\Gamma$ onto a
horizontal subspace of $\tilde\Gamma.$

The generalized system of imprimitivity 
$(\tilde V, \tilde E)$ specified by 
$(\tilde P,\mbox{U}(r), \tilde\Gamma,$id),  with 
id:$\mbox{U}(r)\rightarrow \mbox{U}(r)$ being the
identity mapping (fundamental representation), is equivalent
to $(V,E).$ We describe the corresponding unitary mapping.
If $\psi \in \cal H,$ there is a unique vector 
$\tilde\psi \in \tilde{\cal H}$ such that 
$ \tilde\psi (f(x))=\psi (x).$ Indeed, if $f(x)=f(x'),$ then
$x'=xg$ and $L_g=e,$ hence $\psi (x)=\psi (x')$ for each
$\psi \in \cal H.$ The mapping defined in this way is
unitary and transforms $E$ in $\tilde E$ since
$\tilde\pi (f(x))=\pi(x)$; $V$ is transformed in $\tilde V$
since $f$ preserves the connection.

We can again associate a vector bundle $\tilde F$ to the
principal bundle $\tilde P$ using the fundamental
representation id. Both $\tilde F$ and $F$ have the same
base space and the same typical fibre. In fact, $F$
and $\tilde F$ can be identified by the mapping
$$
W:F \rightarrow \tilde F :[x;\xi]\mapsto [f(x);\xi],$$
where the homomorphism $f:P \rightarrow \tilde P$ was described above.
$W$ is well--defined since 
$$[f(xg);L_g^{-1}\xi]=[f(x)L_g;L_g^{-1}
\xi]=[f(x);\xi].$$ 
$W$ is surjective because $(\tilde
x;\xi)\sim (\tilde x a^{-1};a\xi)=(f(x);a\xi)$ for each
$\tilde x =[x;a]\in \tilde P.$ $W$ is injective because, for
each $u\in M,$ the induced
mapping $W_u:F_u \rightarrow \tilde F_u$ is unitary.

There is again a Hermitian covariant derivative
$\tilde\nabla$ in $\tilde F$ which corresponds to the
Hermitian connection $\tilde\Gamma$ in $\tilde P$. We  can
briefly say that the following diagram commutes:

\begin{picture}(200,80)
\put(100,60){$P(M,G),\Gamma$}
\put(165,65){\vector(1,0){50}}
\put(182,70){$_{f}$}
\put(220,60){$\tilde{P}(M,\mbox{U}(r)),\tilde\Gamma$}
\put(127,57){\vector(2,-1){50}}
\put(127,40){$_{(L)}$}
\put(253,57){\vector(-2,-1){50}}
\put(243,40){$_{(id)}$}
\put(160,20){$F(M,\bbbc^r),\nabla$}
\end{picture}

\noindent
For the corresponding generalized systems of imprimitivity
we find the following commuting diagram:

\begin{picture}(200,80)
\put(110,60){$(V,E)$}
\put(155,65){\vector(1,0){60}}
\put(235,60){$(\tilde V,\tilde E)$}
\put(127,57){\vector(2,-1){50}}
\put(253,57){\vector(-2,-1){50}}
\put(175,20){$(\bar V,\bar E)$}
\end{picture}

\noindent
All these generalized systems of imprimitivity are mutually
unitarily equivalent.

\subsection{The Case $G=\rm{U}(r)$}

With the help of the identical representation id of U$(r)$
in $\bbbc^r$ we can associate a vector bundle $F(M,\bbbc^r)$
to every principal bundle $P(M,\mbox{U}(r)).$ To each $x \in P$
there corresponds the unitary mapping $\hat x$ which we
henceforth denote by the same letter $x$, namely 
$x:\bbbc^r \rightarrow F_{\pi(x)}:\xi \mapsto [x;\xi]$.
Conversely, let $w:\bbbc^r \rightarrow F_{\pi(x)}$ be 
a unitary mapping. Then $w$ necessarily
has the form $\xi \mapsto [x;\xi^w],$ where $\xi \mapsto \xi^w$ is a
unitary mapping $\bbbc^r \rightarrow \bbbc^r.$ Hence there exists 
$a \in \mbox{U}(r)$ such that $w\xi=[x;a\xi]=[xa;\xi]=xa\xi$. 
Every unitary mapping $\bbbc^r \rightarrow F_u$ which is represented 
by unitary mapping $xa : \bbbc^r \rightarrow F_u$ 
is equal to the composition $x \circ a$ since $[xa;\xi ]=[x;a\xi ]$. 
Further, two points $x, x'\in  P$
coincide as unitary mappings if and only if $x=x'.$ So we can return
from the vector bundle $F(M,\bbbc^r)$ back to the principal bundle
$P(M,\mbox{U}(r))$. The fibre $P_u$ over $u \in M$ consists of unitary
mappings $\bbbc^r \rightarrow F_u $ and the structure group 
$G=\mbox{U}(r)$ acts on $P$ by composition $a:x \mapsto x\circ a$.

{\bf Definition 4.1.} Two principal bundles $P(M,G)$, $P'(M,G)$
 over the same base space and with the same structure group
 are said to be {\it isomorphic} if there exists 
a diffeomorphism $f:P \rightarrow P'$ fulfilling
\begin{enumerate}
\item $\pi '(f(x))=\pi (x),$
\item $f(x\circ a )=f(x)a $ for all $a \in G.$
\end{enumerate}
Two Hermitian vector bundles $F(M,\bbbc^r)$, $F'(M,\bbbc^r)$
 are said to be {\it isomorphic} if there exists a diffeomorphism $W:F\rightarrow F'$
such that the restrictions $ W_u:F_u \rightarrow F'_u$ are unitary
mappings for all $u \in M.$

{\bf Lemma 4.1.} Two principal bundles both with structure group
U$(r)$ are isomorphic if and only if the corresponding
associated vector bundles (with typical fibres $\bbbc^r$) are isomorphic.

{\bf Proof.} If $f:P \rightarrow P'$ is an isomorphism, then
$W:F \rightarrow  F':[x;\xi]\mapsto [f(x);\xi]$ is an
isomorphism of the associated vector bundles. Conversely,
let $x \in P$ be a unitary mapping $ \bbbc^r \rightarrow  F_u;$ then $W_u \circ x$ is
unitary mapping $\bbbc^r \rightarrow F'_u$ and there exists
a unique $ x'\in P'_u$ such that $x'=W_u\circ x.$ The
mapping $ P \rightarrow P':x \mapsto W_{\pi(x)}\circ x$ is
the desired isomorphism. $\Box$

As already shown, a connection in a principal bundle can be
carried over to the associated vector bundle. If
$G=\mbox{U}(r)$, $L=$id, this procedure can be inverted. Let us take
a Hermitian connection in the vector bundle $F(M,\bbbc^r)$
associated with principal bundle $P(M,\mbox{U}(r)) $.
The connection relates a family of unitary mappings
 $C_t:F_{u_0}
\rightarrow F_{u_t}$ to every piecewise smooth curve $u_t$
on $M.$ If $x_0 \in P$, $\pi (x_0)=u_0,$ then 
$x_t=C_t \circ x_0$ will be the lift in $P$ of the curve
$u_t$ with the initial point $x_0$. We have
$(C_t \circ x_0)a=C_t \circ (x_0 a).$ This lifting prescription determines a unique connection $\Gamma$ in the
principal bundle $P.$ This correspondence between the
connections in $P$ and the Hermitian connections in $F$ is
one--to--one.

{\bf Definition 4.2.} Connections $\Gamma$, $\Gamma'$ in
 principal bundles $P(M,G)$, $P'(M,G),$ respectively, are
 said to be {\it isomorphic}, if there exists an isomorphism
 $F:P \rightarrow P'$ which maps connection $\Gamma$ in
 connection $\Gamma'.$ Hermitian covariant derivatives
 $\nabla$, $\nabla'$ in Hermitian vector bundles
$ F(M,\bbbc^r)$, $F'(M,\bbbc^r),$ respectively, are said to
 be {\it isomorphic} if there exists an isomorphism 
$W:F \rightarrow F'$ such that $\nabla'=W\nabla W^{-1};$
more precisely,
$$
\nabla'_X \sigma (u)=W_u(\nabla_X W_u^{-1}\sigma (u)).$$

{\bf Lemma 4.2.} Connections $\Gamma,\Gamma'$ in principal bundles
$ P(M,\mbox{U}(r))$, $P'(M,\mbox{U}(r)),$ respectively, 
are isomorphic if and only if the corresponding 
Hermitian covariant derivatives in the associated vector 
bundles (with typical fibres $\bbbc^r$) are isomorphic.

{\bf Proof.} According to Lemma 4.1, to every isomorphism $f:P \rightarrow P'$
there exists an isomorphism $W:F \rightarrow F'$ fulfilling
$f(x)=W_{\pi(x)}\circ x,$ and conversely. If $f$ transforms $\Gamma$ in
$\Gamma'$ and if $x_t$ is the lift in $P$ of a curve $u_t$ with starting
point $x_0,$ then $x'_t=f(x_t)$ is the lift in $P'$ of the same curve
with starting point $f(x_0).$ We have : if $C_t=x_t \circ
x_0^{-1}:F_{u_0}\rightarrow F_{u_t},$ then $C'_t=x'_t\circ
x_0^{,-1}=W_{u_t}C_t W_{u_0}^{-1}.$ Using the last relation we obtain
$\nabla '=W\nabla W^{-1}.$ Conversely, the relation $\nabla'=W\nabla
W^{-1}$ implies $C'_t=W_{u_t}C_t W_{u_0}^{-1}$ and so
$x'_t=W_{\pi(x_t)}\circ x_t.$ this means that the isomorphism
$f:x \mapsto W_{\pi(x)}\circ x$ preserves the connection. $\Box$

\subsection{Unitary equivalence of generalized systems of
im\-pri\-mi\-ti\-vi\-ty}
\par {\bf Theorem 4.2.} Two generalized systems of
 imprimitivity $(V_j, E_j)$ specified by quadruples $(P_j, G_j, \Gamma_J, L_j),
\quad j=1,2$, are unitarily equivalent, if and only if the
corresponding covariant derivatives in the associated vector
bundles are isomorphic.

{\bf Proof.} First we have the necessary condition that the
Hilbert spaces of representations $L_1,\; L_2$ should have
the same dimension, say $r$. According to the results of
Sect. 4.3 we can equivalently  take the generalized systems
of imprimitivity $(\bar V_j, \bar E_j), \quad j=1,2$,
constructed in the associated vector bundles and investigate
their unitary equivalence.

Thus consider a unitary mapping 
$W: \bar {\cal H}_1 \to \bar {\cal H}_2$ relating the two
generalized systems of imprimitivity. Let
$U(F_1,F_2)$ denote the fibre bundle over $M$ with each
fibre over $u\in M$ consisting of unitary mappings $F_{1u}
\to F_{2u}$. From the equality 
$\bar E_2(S) = W\bar E_1(S)W^{-1}$ 
it follows that $W$ is induced by a 
measurable section $u \mapsto W(u)$ in the bundle
$U(F_1,F_2)$. From the equality $\bar{V}^{\phi}_{2}(t) =
W\bar{V}^{\phi}_{1}(t)W^{-1}$ and from relation (4.6) one
deduces that
$$
C_2(\phi_{t} u) = W(\phi_{t} u)C_1(\phi_{t} u)
W^{-1}(u) \eqno(4.8)$$
holds for all $\phi \in D(M)$, all $t\in \bbbr$ and for
almost all (depending on $t$) $u\in M$. 
Here $C_j(\phi_{t} u)$, $j=1,2$, are unitary
mappings corresponding to the curve $u_t=\phi_t u$. 

We shall show that the measurable section $u \mapsto W(u)$,
after proper redefinition on a set of measure zero, is
smooth. It is sufficient to verify this assertion locally,
i.e. to investigate the case $M=\bbbr^n$. 

Let us consider all constant vector fields on
$\bbbr^n$, the corresponding flows and $t=1$. Thus we have for
each $v \in \bbbr^n$ and almost all $u \in \bbbr^n$ 
the equality
$$
W(u+v) = C_2(u+v)W(u)C_1^{-1}(u+v), \eqno(4.9)$$
where $C_j(w): F_{ju} \to F_{jw}$, $j=1,2$.
It follows from Fubini's Theorem that, for almost all $u$,
the equality (4.9) holds true for almost all $v$. 
We need one such $u$. Then we can say that for almost all
$w \in \bbbr^b$ we have
$$
W(w) = C_{2}(w) W(u) C_{1}^{-1}(w).$$
Here the right--hand side depends differentiably
on $w$ and this proves our assertion that the section
$W$ can be defined in smooth manner. From the identity (4.8)
$$
\nabla_X^2 = W\nabla_{X}^{1}W^{-1} \quad \mbox{ for all} 
\quad X\in {\cal X}(M)       \eqno(4.10)$$
now follows immediately.

The converse part of the proof is easier. If covariant
derivatives $\nabla^1$, $\nabla^2$ are isomorphic, there
exists a smooth section $W$ in $U(F_1,F_2)$ which defines a
unitary maping $\bar {\cal H}_1 \to \bar {\cal H}_2 :
 \sigma(u) \mapsto W(u)\sigma(u)$. This unitary mapping
 carries the generalized system of imprimitivity 
$(\bar V_1,\bar E_1)$ over to $(\bar V_2,
\bar E_2)$. Then it suffices to notice that (4.10) implies (4.8). $\Box$

{\bf Corollary 4.1.} Two generalized systems of
 imprimitivity specified by quadruples 
 $(P_j, \mbox{U}(r), \Gamma_j, id)$, $j=1,2$, 
 are equivalent if and only if
 the connections $\Gamma_1$, $\Gamma_2$ are isomorphic.

{\bf Notation.} Let us consider a principal bundle $P(M,G)$ with
connection $ \Gamma$, and fix a point $x_0 \in P, \pi (x_0)=u_0$.
Then, for every piecewise smooth closed curve  $\tau$ on $M$
with base point $u_0$ there exists a unique element $g \in G $
such that $x_0 $ is a starting point and $x_0g$ the end point of
the lift of curve $\tau$. We denote this element by $c(\tau)$.

{\bf Theorem 4.3.} Let $(V_j,E_j)$ be generalized systems of
imprimitivity specified by quadruples $(P_j,G_j,\Gamma_j,L_j)$,
$j=1,2$, and let points $x_1 \in P_1$, $x_2 \in P_2$, $u_0\in
M$ satisfy $\pi_1(x_1) = \pi_2(x_2)=u_0$. The generalized systems
of imprimitivity  $(V_1,E_1)$, $(V_2,E_2)$ are equivalent if and
only if there exists a unitary mapping  $W : {\cal H}^{L_1} \to
{\cal H}^{L_2}$ such that for all piecewise smooth closed
curves $\tau$ on $M$ with the base point $u_0$,
$$L_2(c_2(\tau))=WL_1(c_1(\tau))W^{-1} \eqno(4.11)$$
is valid.

{\bf Proof.} We start with two remarks.
\begin{enumerate}
\item Taking $x'=xa$ instead of $x, \ \pi(x)=\pi(x')=u_0$, then
$c'(\tau)=a^{-1}c(\tau)a$.
\item According to Sect. 4.3, to a principal bundle $P(M,G)$
with connection $\Gamma$ we can associate a principal bundle
$P(M,\mbox{U}(r))$ with connection $\tilde\Gamma$; there also exists a
homomorphism $f_0: P \to \tilde P$ preserving the connection and
fulfilling $f_0(xg)=f_0(x)L_g$. If $\tilde x = f_0(x), \
\tilde\pi(\tilde x) = \pi(x) = u_0$, then $\tilde c(\tau) =
L(c(\tau))$. The systems of imprimitivity for these two principal
bundles are equivalent.
\end{enumerate}

It follows from these two remarks that the choice of a point $x
\in P,\ \pi(x)=u_0$, plays no role and, moreover, we can restrict
our considerations to the case  $G_1 = G_2 = \mbox{U}(r)$,
$L_1 =L_2 = $id. We are thus considering principal bundles 
$P_j(M,\mbox{U}(r))$ with
connections $\Gamma_j,\ j=1,2$. It suffices to show that
condition (4.11) is valid if and only if connections $\Gamma_1,\
\Gamma_2$ are isomorphic.

If $f: P_1 \to P_2$ is an isomorphism preserving the connection
and if we choose $x_2 = f(x)$, then $c_1(\tau)= c_2(\tau)$ holds
for every closed curve $\tau$ with base point $u_0$.

Conversely, let $c_2(\tau)= bc_1(\tau)b^{-1}$ for some $b \in
U(r)$. After having substituted $x_{1}b$ for $x_1$ we can suppose
$c_1(\tau)= c_2(\tau)$. We define a partial mapping $f$ on some
subset of $P_1$ into $P_2$: if $u(t),\ 0\leq t \leq 1$, is a
piecewise smooth curve in $M$ with starting point $u_0$ and if
$x_1(t),\ x_2(t)$ are the lifts of this curve in $P_1,\ P_2$ with
starting points $x_1,\ x_2$, respectively, we put $f(x_1(1))=
x_2(1)$. Let us investigate the case when two curves $u(t),\
v(t),\ 0\leq t \leq 1,\ u(0)= v(0)= u_0$, have the same end
point $u(1)= v(1)$. Let $x_j(t),\ y_j(t),\ j=1,2$ be the lifts of
curves $u(t),\ v(t)$ with starting points $x_1$, $x_2$,
respectively. There exists a unique $a \in G $ such that $y_1(1)=
x_1(1)a$. Let $\tau(t),\ 0\leq t\leq 1$, be a closed curve with
base point $u_0$, coinciding with $u(t)$ for $0\leq t\leq 1$ and
with $v(2-t)$ for $1\leq t\leq 2$. Then we find $c_1(\tau)=
a^{-1}= c_2(\tau)$ and hence $y_2(1)= x_2(1)a$. Thus function $f$
is well--defined. The domain of $f$ consists of all points in
$P_1$ which can be connected with $x_1$ by a horizontal curve.
But this domain can be extended to the whole $P_1$ by the
relation $f(xa)= f(x)a$. In this way $f$ becomes an isomorphism
of principal bundles $P_1 \to P_2$ and, by construction,
preserves the connection. $\Box$

\subsection{Irreducibility of generalized systems of imprimitivity}

Let us consider principal bundle $P(M,G)$ with connection
$\Gamma$. We associate a subgroup $\Phi_x$ of $G$ to each point
$x \in P;\ \Phi_x$ consists of all $g\in G$ such that the points
$x,\ xg$ lie on a common horizontal curve in $P$. The group
$\Phi_x$ is called the {\em holonomy group} of the connection
$\Gamma$ with the base point $x$. It has following properties:
\begin{enumerate}
\item If $x'\in P$ can be connected with $x$ by a horizontal
curve, then $\Phi_{x'}=\Phi_x $;
\item $\Phi_{xb}=b^{-1}\Phi_x b$.
\end{enumerate}

So all holonomy groups are conjugate subgroups in $G$ and we need
not specify the base point. The {\em restricted holonomy group}
$\Phi^0$ is the subgroup of $\Phi$ corresponding to
 horizontal lifts of those closed  curves which are
 homotopic to $0$. The groups $\Phi,\ \Phi^0$ -- being
 subgroups of $G$ -- are topological groups.

{\bf Theorem 4.4}. (\cite{27}, Chap. II.3, \cite{24, GHV}). The
restricted holonomy group $\Phi^0$ is a connected Lie group, and
it coincides with the arcwise connected component of unity in
$\Phi$. Moreover, the quotient $\Phi/\Phi^0$ is finite or
countable.

The holonomy group $\Phi $ itself need not be a Lie group. But
it can be equipped with a new topology which induces the original
topology on $\Phi^0$ and the quotient group $\Phi/\Phi^0$ is
discrete. In this topology one can verify that the inclusion
$\imath: \Phi \hookrightarrow G : g \mapsto g$ is a
homomorphism of Lie groups.

{\bf Definition 4.3.} We say that a structure group $G$ of a
principal bundle $P(M,G)$ is {\em reducible} to a Lie group
$G'$, if there exists a principal bundle $P'(M,G')$ and
 homomorphism $f: P' \to P$ such that
 $f(x'g')=f(x')f_0(g'),\ \pi(f(x'))= \pi'(x')$ 
with $ f_0 :G' \to G $ being an injective homomorphism
of Lie groups. Moreover, if bundles $P,\ P'$ are endowed
 with connections $\Gamma,\ \Gamma'$, respectively, and the
homomorphism $f$ preserves connection, we say that
connection $\Gamma $ is {\em reducible} to connection
$\Gamma'$.

{\bf Theorem 4.5.} Let a generalized system of imprimitivity
$(V,E)$ be specified by a quadruple $(P,G,\Gamma,L)$. If
connection $ \Gamma $ is reducible to a connection $\Gamma'$ on a
principal bundle $P'(M,G')$ and $L^0= L\circ f_0$ denotes the
representation of the Lie group $G'$, then the generalized system
of imprimitivity specified by the quadruple $(P',G',\Gamma',L^0)$
is equivalent to $(V,E)$.

{\bf Proof.} The assertion can be proved by a method completely
analogous to that used at the end of Sect. 4.3. The desired
unitary mapping can be constructed in terms of the injective
homomorphism $f:P' \to P;\ \ \psi'(x')= \psi(f(x'))$.$\Box$

{\bf Theorem 4.6} (\cite{27}, Chap. II.6, \cite{24}).
Let $P(M,G)$ be a principal bundle, $\Phi$ the holonomy group of
connection $\Gamma$ in $P$. Then the structure group $G$ is
reducible to a connection in the reduced principal bundle
$P'(M,\Phi)$, the holonomy group  of which is identical with
$\Phi$.

{\bf Theorem 4.7.}
Let $(V,E)$ be a generalized system of imprimitivity specified by
a quadruple $(P,G,\Gamma, L),\ \Phi$ be the holonomy group of
connection $\Gamma, \ L^0$ be the restriction of representation
$L$ to the subgroup $\Phi$. Then  the commuting algebras ${\cal
C}(V,E),\ {\cal C}(L^0)$ are isomorphic.

{\bf Proof.} In view of Theorems 4.5 and 4.6 it suffices to
consider the case $G=\Phi$. Let $\tilde A$ be a bounded operator
${\cal H} \to {\cal H}$. If $\tilde A$ commutes with all $E(S),\
S\in {\cal B}(M)$, then, according to Theorem 3.4, it is of the
form $\psi(x) \mapsto A(x)\psi(x)$, where $x \mapsto A(x)$ is
some measurable mapping from $P$ into the space of operators in
$\bbbc^r$ such that $A(xg)= L_g^{-1}A(x)L_g$ is true on almost
all fibres. Moreover, if $\tilde A$ commutes with all
$V^{\phi}(t)$, then $A(x)=V^{\phi}(t)A(x)V^{\phi}(t)^{-1}
=A(\tilde\phi_t\cdot x)$ holds almost everywhere. We shall
show that function $x \mapsto A(x)$ can be considered smooth
after a redefinition on a set of measure zero. To show this local
property, we can consider $ M=\bbbr^n,\ P(M,G)= \bbbr^n\times G$. Then
for vectors $v\in S^{n-1}\subset \bbbr^n$ the mapping $y \mapsto
A(y)$ remains constant (almost everywhere) on horizontal lifts of
straight lines in $\bbbr^n$ with directions $v$ and passing through
$u$ for almost all $u$. More precisely, if $\phi_t^v(u)=
 u+ tv$, then for all $(t,v) \in \bbbr \times S^{n-1} $ and
 for almost all $u\in \bbbr^n$  (depending on $t$ and
$v$) the equality
$$
A(u;g) = A(\tilde\phi_t^v(u;g)) \eqno(4.12)$$
is valid for each $g\in G$. Using the
Fubini Theorem we find that for almost all $u\in \bbbr^n$, (4.12)
holds for all $g\in G$ and almost all $(t,v)\in \bbbr\times S^{n-1}$.
Let us fix $u$ with this property; we can assume $u=0$. Then we
construct an auxiliary section $\sigma$ in $P$. For $w\in \bbbr^n$ we
lift the curve $u(t) = tw,\ 0\leq t\leq 1$, in the given
connection, choosing $(0;e)$ for the starting point, and we put
$\sigma(w)$ to be equal to the final point of the lifted curve.
Let $A =A(0;e)$. Then for almost all $w\in \bbbr^n$ and all $g\in G,\
A(\sigma(w)g) = L_g^{-1}AL_g$ holds. This proves our assertion
that $x \mapsto A(x)$ can be considered smooth. Namely, the
mapping $(w,g) \mapsto \sigma(w)g$ is an automorphism of the
principal bundle $\bbbr^n\times G$.

So let us suppose that function $x \mapsto A(x)$ is smooth and
again use the equation $A(x) = A(\tilde\phi_t\cdot x)$,
now being valid for all $t,\ x$. We find that the linear mapping
$\d A_x : T_xP \to \bbbc^{r,r}$, if restricted to the horizontal
subspace, is zero. Hence $A(x)$ is constant on horizontal curves
in $P$. Since the holonomy group coincides with the structure
group, arbitrary two points in $P$  can be connected by a
horizontal curve. Thus $A(x) = A$ for all $x\in P$ and we have $A
= A(x)= A(xg)= L_g^{-1}AL_g$. In this way we have associated
a unique operator $A\in {\cal C}(L^0)$ to each $\tilde A \in
 {\cal C}(V,E)$.

Conversely, one can relate a unique $\tilde A \in {\cal C}(V,E)$
to each $A\in {\cal C}(L^0)$ by means of the relation $\tilde A
\psi(x)= A\cdot \psi(x)$. The one--to--one correspondence $\tilde
A \leftrightarrow A$ is the desired isomorphism. $\Box$

{\bf Corollary 4.2.} $(V,E)$ is irreducible if and only if $L^0$
is irreducible.

\subsection{Quantum Borel kinematics with vanishing 
           external field}
Let $\cal G$ denote the Lie algebra of the group $G$, u$(r)$
the Lie algebra of the group U$(r)$ (consisting of 
skew--Hermitian
$(r \times r)$--matrices), and $\Omega$ the curvature 
2--form of connection $\Gamma$. $\Omega$ is a 
2--form on $P$
taking values in $\cal G$ such that $R_a^*\Omega = 
{\rm ad}(a^{-1})\Omega$ holds for all $a\in G$, where 
$R_a : P \to P : x \mapsto xa$. By composition with the
representation $L'$ of $\cal G$ we obtain a
2--form  $L'\circ \Omega$ taking values in u$(r)$. 
Under the homomorphism of principal bundles 
$f : P\to \tilde P,\ \tilde P= \tilde
P(M,\mbox{U}(r))$  (see Sect. 4.3) $L'\circ \Omega$ is
 mapped into the curvature form $\tilde \Omega$ of the
 connection $\tilde\Gamma$. 

For a vector bundle $F$ associated to $P$ let
End $F$ denote a vector bundle over $M$, with fibres 
(over $u \in M$) consisting of linear endomorphisms
of fibres (from $F_u$ into $F_u$).
2--forms $w$ on $P$ taking values in $u(r)$ and satisfying
$R_a^*w = {\rm ad}(a^{-1})w,\ a\in \mbox{U}(r)$, are in 
one--to--one
correspondence with 2--forms $K$ on $M$ taking skew--adjoint
 values in the
space of sections $\mbox{Sec}(\mbox{End} F)$; the correspondence is expressed
by  the relation 
$$K_u(X,Y)= x\circ w_x(X^*,Y^*)\circ x^{-1},$$
where $\tilde\pi(x) =u$ and $X^*,\ Y^*$ are horizontal lifts
of $X,\ Y$ with respect to the connection $\tilde\Gamma$. In
this way to $\tilde\Omega$ a 2--form $R$ is related and
$$
R(X,Y) = \left[\nabla_{X},\nabla_Y\right]
-\nabla_{\left[X,Y\right]} \eqno{(4.13)}$$
holds (\cite{27}, Chap. III.5). $R$ is the {\it curvature}
 of the covariant derivative $\nabla$. 
A simple calculation using (4.7) then leads to
$$
\left[P(X),P(Y)\right] = -i\hbar P(\left[X,Y\right])
    + (-i\hbar)^{2} R(X,Y),
$$ 
since $P \approx \tilde{P}$.

If $R=0$ (this is
 true if and only if $L'\circ \Omega = 0$), we say that the
 {\em external field vanishes} on the manifold. 
We see that the field vanishes if and only if 
$$
\left[P(X),P(Y)\right] = -i\hbar P(\left[X,Y\right]),$$ 
see Sect. 3.2, eq. (3.3).

We recall the mapping $c : \tau \mapsto c(\tau)$ introduced in
Sect. 4.5. If the field vanishes on the manifold, then the value
$L(c(\tau))\in U(r)$ depends only on the homotopy class of the
curve $\tau$. In this manner we obtain a representation $L^c$ of
the fundamental group $\pi_1(M)$ of manifold $M$. Let
$(M^c,\pi^c,M;\pi_1(M))$ be the universal covering of $M$. Since
$\dim M^c = \dim M$, there exists exactly one flat connection
$\Gamma^c$ on the principal bundle $(M^c,\pi^c,M;\pi_1(M))$.
Moreover, according to Theorem 4.3 (Eq. (4.11)) the generalized
system of imprimitivity $(V^c,E^c)$ specified by the quadruple
$(M^c,\pi^c,\Gamma^c,L^c)$ is equivalent to $(V,E)$. Conversely,
since the connection $\Gamma^c$ is flat, the field on  $M$ will
vanish for every generalized system of imprimitivity $(V^c,E^c)$,
no matter which representation $L^c$ of $\pi_1(M)$ is
 chosen. So we arrive at a canonical form for generalized
 systems of imprimitivity (or quantum Borel kinematics)
with vanishing field (with flat connection).
This form was already studied in detail (see \cite{2,3} and
 references therein).
The results of Sections 4.5 and 4.6 imply in this case:

{\bf Theorem 4.8.} Two generalized systems of imprimitivity
 (or two quantum Borel kinematics) with vanishing field
$(V_j^c,E_j^c),\ j=1,2$, are equivalent if and only if the
representations $L_1^c$, $L_2^c$ are equivalent. The
 commuting algebras ${\cal C}(V^c, E^c),\ {\cal C}(L^c)$ are
 isomorphic.

\section{Quantum Borel kinematics: classification}

\subsection{Classification of generalized systems
       of imprimitivity via cocycles}
The  generalized systems of imprimitivity described by
(4.1), (4.2) are not of the most general form. Starting from
a characterization of Mackey's systems of imprimitivity
in terms of cocycles (\cite{30}, Theorem. 9.11), 
the following theorem was proved in \cite{Natter}:

{\bf Theorem 5.1.} Any $r$--homogeneous
generalized system of imprimitivity on $M$
is unitarily equivalent to a canonical one $({V},{E})$, with
$ {\cal H} = L^{2}(M,\bbbc^{r},\mu)$
for some smooth measure $\mu$ on $M$,
$$ ({E}(S)\psi)(u) = \chi_{S}(u) \psi (u)$$
for all $\psi \in {\cal H}$ and $S \in {\cal B}(M)$,
and
$$
[{V}^{\phi^{X}}(t)\psi ](u) = \xi^{X}(t,
\phi_{-t}^{X}(u))\sqrt{\frac{\d \mu}{\d (\mu \circ
 \phi_{t}^{X})}(\phi_{-t}^{X}(u))} \psi(\phi_{-t}^{X}(u))
\eqno{(5.1)}$$
for all $\psi \in {\cal H}$ and all $X \in
{\cal X}_{c}(M)$. Equivalence classes of $r$--homogeneous
generalized systems of imprimitivity are in one--to--one
correspondence with equivalence classes of cocycles
$[\xi^{X}]$.

Here $\xi^{X}$ is a {\it cocycle} of $\bbbr$ relative to 
the Lebesgue measure class on $M$ with values in U($r$),
i.e. a Borel measurable map
$  \xi^{X} : \bbbr \times M \mapsto \rm{U}(r) $
with
$$\xi^{X} (0, u)  =  1, $$
$$\xi^{X} (s + t, u)  =  \xi^{X}(s, \phi^{X}_{t}(u))
\xi^{X}(t,u)
$$
for almost all $u \in M$ and almost all $s,t \in \bbbr.$
Two cocycles $\xi^{X}_j$, $j= 1,2$, are called {\it
 equivalent (cohomologous)}, if there is a Borel function
$\zeta : \bbbr \mapsto \rm{U}(r)$, such that for all
$X \in {\cal X}_{c}(M)$ and $t \in \bbbr$, $u \in M$
$$
\xi^{X}_{2} (t, u)  = \zeta(\phi^{X}_{t}(u)) \xi^{X}_{1}
  (t,u)\zeta(u)^{-1}. $$

Unfortunately, the classification given in Theorem 5.1
is not easy to handle, since the calculation of cocycles
is rather tedious. To be more specific, one has to impose
further conditions on the operators under consideration.

\subsection{Differentiable quantum Borel kinematics}
To gain further insight into the structure of 
the shift operators (5.1), one can perform
a formal calculation.
In particular, assume for the moment that
the cocycles of the representation are {\it smooth} maps
from $\bbbr \times M$ into U($r$). Then,
by formal differentiation of (5.1) with respect to $t$
at $t=0$, an expression for the generalized
 momentum operator $P(X)$ is obtained,
$$P(X) \psi = -i\hbar [X \psi +  
\frac{1}{2} \rm{div}_{\mu} (X)\cdot \psi ] - 
i\hbar \rho (X) \cdot \psi, 
\eqno{(5.2)}$$
where $\psi \in C^{\infty}_{0} (M,\bbbc^r)$ and
$$ \rho (X)(u) :=  \frac{\d}{\d t}
  \xi^{X}(t, \phi^{X}_{-t} (u))|_{t=0}.
$$
The first two terms on the right hand side of (5.2) 
are linear in the vector field $X \in {\cal X}(M)$.
Though the set of complete vector fields ${\cal X}_{c}(M)$
is not a linear space --- the sum of two complete vector
fields may not be complete --- it contains the ``large''
linear subset ${\cal X}_{0}(M)$ of vector fields 
with compact support for which one
can demand linearity (or demand `partial' linearity 
at least for all {\it complete}
linear combinations of complete vector fields), cf.
Theorem 3.1). Thus
as a first additional assumption on $P(X)$ we require (3.2),
i.e. $\rho (X)$ to be linear in that case.

Using the formal expression (5.2), the commutator of
$P(X)$ and $Q(f)$ is obviously obtained again in the form
 (3.5). Finally, for the commutator $[P(X), P(Y)]$ we obtain
$$
[P(X), P(Y)] = -i\hbar P([X,Y]) -\hbar^{2} R(X,Y),
\eqno{(5.3)}$$
where
$$ R (X,Y) =  [\rho(X),\rho(Y)] + 
  X \rho (Y) - Y \rho (X) - \rho([X,Y]).
\eqno{(5.4)}$$
If $ \rho$ were a localized connection 1--form,
(5.4) would represent the local definition of 
a curvature 2--form $ R$ of a
$\bbbc^{r}$--bundle over $M$. The Jacobi identity
for generalized momenta would then give us precisely
the Bianchi identity $D R = 0$, where $D$ is
the covariant differential defined by the connection.

In order to arrive at this point we had to assume
differentiability of the shift operators (5.1) and of
the functions in the domain of momentum operators.
Now there are different ways of defining differentiable
structures and thus differentiability on the set
$M \times \bbbc^r$; for a discussion of this point we
refer to \cite{3,Drees}. On the other hand, we have
already interpreted $R$ as a curvature 2--form
that is in general related to a connection on a
$\bbbc^r$--bundle over $M$. This line of reasoning 
leads us to the following definition \cite{Drees,Natter}:

{\bf Definition 5.1.} Let $M$ be a  differentiable manifold,  
$(V,E)$ an $r$-homogeneous generalized system of
imprimitivity on $M$ for $r= 1,2, \ldots$,
and $R$ a differential 2--form on $M$ with values
in Hermitian operators on $\bbbc^r$. Let $Q(f)$ and
$P(X)$ denote the corresponding generalized position
and momentum operators, respectively. Then:
\begin{description}
\item  1) The quadruple $({\cal H}, Q,P,R )$ is called
 an {\it ($ R$--compatible) quantum Borel  
$r$--kinematics} (QBK$^r$), if $P$ is (partially) linear, satisfies (5.3), (5.4), and the common
invariant domain ${\cal D}$ of $Q(f)$'s and $P(X)$'s for 
$f \in C^{\infty} (M)$ and $X \in {\cal X}_{c}(M)$ is dense
in  ${\cal H}$. It is {\it local},
if $P$ is local, and {\it elementary}, if $r=1$.
\item  2)  A quantum Borel kinematics is called 
{\it differentiable}, if it is equivalent to
$({\cal H}, Q,P,R )$ of {\it standard form},
constructed from the following ingredients:
\begin{itemize} 
\item[1.] {Lebesgue measure $\mu$ on $M$;} 
\item[2.] {Hermitian vector bundle $F(M;\bbbc^r)$ over $M$
 with fibres diffeomorphic to
$\bbbc^r$ equipped with Hermitian inner product $<.,.>\,$;}
\item[3.] {A 2--form $R$   with skew--adjoint values
in the endomorphism bundle 
$ \mbox{End}\, F = F \otimes F^*$;}
\item[4.] {The Hilbert space ${\cal H}$ is realized
as the Hilbert space $L^{2}(F,<.,.>,\mu)$ of sections of
$F$, i.e.\ (measurable) mappings $\sigma\colon M\to F$ such
 that $\bar{\pi}\circ\sigma= {\rm id}_M$ and with
finite norm induced by the inner product 
$$(\sigma ,\tau)=\int_{M}^{}
<\sigma (u),\tau (u)>\, {\rm d}\mu (u) \, ;
$$ }
\item[5.] {The common invariant domain ${\cal D}$
for $Q$, $P$ contains the set $\rm{Sec}_{0}\, F$ of 
smooth sections of $F$ with compact support and
 $P(X)\rm{Sec}_{0}\,F \subset \rm{Sec}_{0}\, F$;}
\item[6.] {The position operators $Q(f)$ have the usual form
of the Schr\"{o}dinger representation
$$
Q(f)\sigma=f\cdot\sigma,\qquad \forall 
f\in C^{\infty}(M,\bbbr), \quad \sigma \in \rm{Sec}_{0}\,F
\,. $$ }
\end{itemize} 
\end{description}

\subsection{Canonical representation of differentiable
QBK$^r$}
For {\it local differentiable quantum Borel kinematics}
the formal calculations can be made precise. According to
Sect. 5.2 it only remains to derive
the representation of generalized momenta $P(X)$ in
a standard form. This is the content of 

{\bf Theorem 5.2} (\cite{3,Drees}).
Let  $({\cal H}, Q,P,R )$ be a local differentiable
quantum Borel kinematics on $M$ in a standard form.
Then there is
\begin{itemize}
\item[1.] {a Hermitian connection $\nabla$ 
with curvature $R$ on $F$, i.e. 
a connection compatible with the inner product 
$$X<\sigma,\tau>=<\nabla_{X}\sigma,\tau>+<\sigma,
 \nabla_{X}\tau>\, ,$$}
\item[2.] {a covariantly constant self--adjoint section
$\Phi$ of End $F = F \otimes F^*$, the bundle
of endomorphisms of $F$,} 
\end{itemize} 
such that for all $X \in {\cal X}_{c}(M)$ and all
$\sigma \in \rm{Sec}_{0}\, F$
$$
P(X)\sigma=-i\hbar\nabla_{X}\sigma + ( -{i\hbar \over 2}I
+ \Phi)Q(\mbox{div}_{\nu}X)\sigma, \quad
\sigma\in {\cal D}. \eqno(5.5)$$ 
Moreover, $R$ is a curvature 2--form on $F$
satisfying the Bianchi identity
$$ D R = 0, $$
where $D$ denotes the covariant differential defined
by the connection $\nabla$.

The canonical form of generalized momenta (5.2) shows
that by imposing $R$--compatibility the quantum
system on $M$ is influenced by an external classical
field on $M$. In general, this field is
defined by a curvature 2--form on $F$, and thus
 a curvature $\tilde{\Omega}$ on the associated   
U$(r)$--principal bundle $\tilde{P}(M,\mbox{U}(r)$. 
We could think of this   
$\tilde{\Omega}$ as a classical gauge  (Yang--Mills) field.
The simplest example was provided in Sect. 4.1 for
$M = \bbbr^3$, $r=1$, where the 2--form $e\beta$ was
interpreted as a coupling constant times a magnetostatic
field on $\bbbr^3$, with the Bianchi identity corresponding
to the Maxwell equation $\rm{div}\vec{B} = 0$.
Up to a coupling constant the connection $\nabla$ 
generalizes the notion of a vector potential.
For $M = \bbbr^3$ the global connection form $\alpha$
corresponds to the vector potential $\vec{A}$ and the
2--form $\beta = \d \alpha$ to the magnetic field
$\vec{B} = \rm{rot} \vec{A}$ of Maxwell's theory.

\subsection{Classification of differentiable QBK$^r$'s}
The canonical form given in Theorem 5.2 indicates that
a classification of differentiable quantum Borel kinematics
amounts to a classification of Hermitian $\bbbc^r$--bundles
with connection over $M$ {\it and} covariantly constant 
self--adjoint sections of the corresponding endomorphism
bundle.\footnote{The zero section of End $F$ always
exists and is covariantly constant and self--adjoint.} 
This is the content of

{\bf Theorem 5.3}. Two local differentiable
quantum Borel kinematics  $({\cal H}_j, Q_j,P_j,R_j )$,
$j = 1,2$, in canonical form of Theorem 5.2 are equivalent,
if and only if there is a strong, unitary, and connection
(and thus curvature) preserving bundle isomorphism
$I: F_1 \rightarrow F_2 $ mapping $\Phi_j$ into each
other, i.e.
$$ \nabla_2 = I \circ \nabla_1 \circ I^{-1}, \quad
 R_2 = I \circ R_1 \circ I^{-1},  \quad
 \Phi_2 = I \circ \Phi_1 \circ I^{-1}.
$$

Unfortunately, there are no general existence and 
classification theorems of Hermitian $\bbbc^r$--bundles with connection. Looking back to Sect. 4, there a rather big
class of local differentiable QBK$^r$'s is constructively
defined, however with $\Phi = 0$. Hence even in these
cases, the additional classification of covariantly constant
self--adjoint sections has to be found as well.
This last problem was solved only in certain special cases
--- elementary quantum Borel kinematics and type 0 or
type U(1) QBK$^r$'s --- described in the following
sections.

\subsection{Classification of elementary 
  differentiable quantum Borel kinematics}
The problem of existence and classification of
{\it elementary}, i.e. $r=1$ local differentiable quantum Borel kinematics in terms of global geometrical properties
(cohomology groups) of the underlying manifold $M$ 
was completely solved \cite{3,Drees}. It is based on
a theorem \cite{Ko,GP} concerning existence and
classification of complex line bundles with hermitian
connection.

{\bf Theorem 5.4} Let $M$ be a connected
differentiable manifold and $B \in \Lambda^{2}(M)$ be 
a closed 2--form on $M$ with $\d B = 0 $. 
Then there exists a complex line bundle
$F$ with hermitian connection $\nabla$ of 
curvature $R = \frac{i}{\hbar} B$ if and only if
$R$ satisfies the integrality condition
$$ \frac{1}{2\pi i} \int_{\sigma}^{} R =
 \frac{1}{2\pi \hbar} \int_{\sigma}^{} B \in
  \bbbz $$
for all closed 2--surfaces $\sigma$ in $M$. 
In terms of cohomology theory, the de Rham class of
$R / (2\pi i) = B / (2\pi \hbar) $ has to be integral,
$$[ \frac{1}{2\pi \hbar} B] \in H^{2}(M,\bbbz).$$

Hence non--isomorphic equivalence
classes of principal bundles over a manifold $M$ with the
structure group U(1) are labeled by elements of the second
cohomology group $H^2(M,\bbbz)$.
The Lie algebra of U(1) coincides with the imaginary axis
$i\bbbr$. Since U(1) is Abelian, the vector bundle End $F
=M\times \bbbc$ is trivial. So the curvature $R$ is a
purely imaginary 2--form on $M$, $R=\Omega$, 
where $\Omega$ is the curvature 2--form
of connection $\Gamma$. If we put
$$\beta=-\frac{i\hbar}{e} R, $$
$\beta$ can be interpreted as the 2--form of 
{\it external magnetic field} on $M$. For an arbitrary 2--cycle
$\sigma$ of the singular homology on $M$, 
$\partial \sigma = 0$, we have 
$$
\exp(\int_{\sigma}\Omega)=1.$$ 
This leads to the {\it Dirac quantization condition}
on the magnetic field
$$
\int_{\sigma} \beta = g,\quad \mbox{with} \quad
          \frac{eg}{2\pi}=n\hbar,$$  
where $n\in \bbbz$.
We may interpret this result that the 2--cycle $\sigma$
--- besides the usual magnetic field satisfying 
$ \int_{\sigma} \beta=0$ ---  encloses a
{\it Dirac magnetic monopole} with quantized magnetic charge $g$.

Furthermore, the various inequivalent choices of 
$(F,<.,.>,\nabla)$ for fixed curvature $R$
are parametrized by
$$H^{1}(M, \rm{U}(1)) = \pi_{1}(M)^*, $$
where $\pi_{1}(M)^* $ denotes the group of characters
of the fundamental group of $M$.

We should emphasize that $H^{1}(M, \rm{U}(1)) =
 \pi_{1}(M)^* $ classifies pairs of hermitian line bundles
$(F,<.,.>)$ {\it and} compatible connections $\nabla$.
This implies that the curvature 2--forms of two equivalent
complex line bundles with hermitian connection are
identical. The classification of complex line bundles
$F$ themselves --- disregarding their connection ---
is given by elements of the \v{C}ech cohomology
\v{\it H}$^{1}(M,\rm{U}(1)) = H^{2}(M,\bbbz)$:
two complex line bundles are equivalent if and only if 
their Chern classes in $H^{2}(M,\bbbz)$ coincide,
i.e. the curvature 2--forms {\it admissible} on these
two bundles are in the same integral de Rham cohomology
class of $H^{2}(M,\bbbz)$.

Theorem  5.4 is the basis of a classification theorem 
that goes back to \cite{3} for the flat ($B=0$) case 
and was extended to the case of external fields
$B$ by \cite{Drees}:

{\bf Theorem 5.5}. The equivalence classes
of elementary local differentiable quantum Borel kinematics
are in one--to--one correspondence to elements of the set
$$ 
 H^{2}(M,\bbbz) \times H^{1}(M, \rm{U}(1)) \times \bbbr. $$

For the proof it remains to classify the inequivalent
choices of covariantly constant self--adjoint sections
$\Phi$ of End $F$. For a line bundle the endomorphism
bundle is actually trivial: As the transition functions
$\varphi_{jk} : U_{j} \cap U_{k} \rightarrow \rm{U}(1)$ of
$F$ commute with complex numbers, the induced transition
functions of End $F = F \otimes F^*$ become
trivial, $z \mapsto \varphi_{jk}z \varphi^{*}_{jk} = z$,
hence End $F = M \times \bbbc$. Thus the sections of
this bundle correspond to complex functions on $M$. 
Furthermore, the induced connection on $M$ is the trivial 
connection on $M \times \bbbc$ given by the Lie derivative.
Thus covariantly constant self--adjoint sections $\Phi$ of End $F$ are real multiples of the identity,
$$ \Phi = \hbar c\; \cdot
 \rm{id}_{\mbox{Sec} F}, \quad c \in \bbbr.$$
Obviously, $c$ is not changed under strong bundle
isomorphisms $I$, so each value of $c \in \bbbr$ for
a given Hermitian line bundle determines an 
inequivalent local differentiable quantum Borel
kinematics. $\Box$

Finally let us note that elementary quantum Borel kinematics
with $c=0$ are, in terms of constructions of Sect. 4,
described by generalized systems of imprimitivity 
$(V,E)$ specified by
the quadruple $(P,\mbox{U}(1),\Gamma,$ id).

\subsection{Classification of quantum Borel kinematics of
 type $0$} 
The whole variety of quantizations could be read off the
formula (5.5). In order to get a more transparent
result we define a QBK$^r$  {\it of type 0} \cite{Mu} by
$$
\Phi=\hbar c.{\rm id}_{\mbox{Sec}\, F},\qquad c\in \bbbr.$$ 
Then we obtain an identical formula for $P(X)$ as in QBK$^1$ 
\cite{4}:
$$
P(X)\sigma=-i\hbar\,\nabla_{X}\sigma+
(-{i\hbar \over 2}+ \hbar c)
(\mbox{div}_{\mu}X)\cdot\sigma,\quad \sigma\in {\cal D}\,.
\eqno(5.6)$$ 
As proved in \cite{Mu}, on every smooth
manifold $M$ there exists a differentiable QBK$^r$ of type
 $0$. Let us note that for $r=1$, the type $0$
 QBK$^1$'s {\it classify all possible Borel quantizations} 
\cite{4}; this is not the case, however, for $r>1$.

Finally, a complete classification of QBK$^r$'s of type $0$
was possible in the case of {\it flat connection} ($R = 0$)
\cite{4, 14, Mu}. It turns out that it is essentially 
the question of the topology of $M$. The
corresponding investigations can be summarized in

 {\bf Theorem 5.6} The set of classes of
unitarily equivalent QBK$^r$'s of type $0$ with flat 
connection on $M$ can be
bijectively mapped onto the set of pairs $({\cal L},c)\,$, 
where $c\in \bbbr$ and ${\cal L}$ denotes 
the isomorphism class of flat $\bbbc^r$-bundles over $M$.

Since there is a one-to-one
correspondence between the isomorphism classes of flat
$\bbbc^r$-bundles over $M$ and flat U$(r)$-principal bundles
over $M$, we can use Milnor's

{\bf Lemma} \cite{Mi}. A U$(r)$-principal bundle over $M$
 admits a flat connection if and only if it is induced from
 the universal covering bundle of $M$ by a homomorphism of
 the fundamental group $\pi_{1}(M)$ into U$(r)$.

Thus, disregarding the real constant $c$, the set of
 inequivalent quantizations of type $0$ with 
$\bbbc^r$--valued wave functions is isomorphic to the set 
$$\mbox{Hom}(\pi_{1}(M), \mbox{U}(r))$$ 
of (the conjugacy classes of) $r$-dimensional unitary
 representations of the fundamental group of $M$.

In the case $r=1$, i.e. of quantizations with 
complex--valued wave functions, the topological part of the
 classification
reduces to $\mbox{Hom}(\pi_{1}(M), \mbox{U}(1))$, i.e. to the set
of one-dimensional unitary representations of $\pi_{1}(M)$
\cite{3,4,Do-To,Mu}.\footnote{This
result was obtained independently also in the Feynman path
integral approach \cite{9} and in geometric quantization
\cite{Ko}.} Since the commutator subgroup $\Gamma(\pi_{1}(M))$
(generated by elements $aba^{-1}b^{-1}$) belongs to the kernels
of all such one-dimensional representations and since the
singular homology group $H_{1}(M,\bbbz)$ is isomorphic to
$\pi_{1}(M)/\Gamma(\pi_{1}(M))$ (the Hurewicz isomorphism),
inequivalent QBK$^1$'s are labeled by elements of the character
group of $H_{1}(M,\bbbz)$. 

Finally, let us describe the general
structure of the Abelian group $H_{1}(M,\bbbz)$ for compact $M$. It
has a decomposition $H_{1}(M,\bbbz)=F\oplus T\,$, where the free
Abelian group $F$ is $F=\bbbz\oplus \cdots \oplus \bbbz$ ($b_1$ terms),
with $b_1$ being the {\it first Betti number} of $M$, and the
torsion Abelian group is $T=\bbbz_{\tau_1}\oplus\cdots \oplus
\bbbz_{\tau_k}$ with $\bbbz_{\tau_i}$ being cyclic groups of orders
$\tau_i$ ({\it torsion coefficients}) such that
$\tau_{i+1}/\tau_{i}= \,$ positive integer. Thus the characters
of $H_{1}(M,\bbbz)$ can be parametrized by $(b_{1}+k)$-tuples
$$[e^{2\pi i\theta_{1}},\cdots ,e^{2\pi i\theta_{b_1}}\;;
\;e^{2\pi im_{1}/\tau_{1}},\cdots ,e^{2\pi im_{k}/\tau_{k}}]$$
with the numbers $\theta_{l}\in [0,1)\,$, $l=1,\dots,b_1\,$, and
$m_{i}=0,1,\dots,\tau_{i}-1\,$, $i=1,\dots,k\,$, classifying
inequivalent quantum Borel 1--kinematics on $M$.

\subsection{Elementary quantum Borel kinematics with
 vanishing external field} 
It is remarkable that elementary quantum Borel kinematics 
with {\it vanishing external magnetic field} 
find application in quantum mechanics 
(Aharonov-Bohm effect \cite{1}).
According to Sect. 4.3 they can be labeled by 
one--dimensional unitary
representations of the fundamental group $\pi_{1}(M)$,
and consequently, as already explained in Sect. 4.9,
inequivalent systems of imprimitivity with vanishing
magnetic field are labeled by elements of the
character group of $H_1(M,\bbbz)$.

Quantum Borel kinematics in the case of 
{\it trivial fibration} $P=M \times \mbox{U}(1)$ was studied in
detail in \cite{8}. In this case a localized connection 
1--form $i(e/\hbar)\alpha$ can be defined globally on the
whole manifold $M$; the closed 1--form $\alpha$ represents the
vector potential of the vanishing magnetic field. 
A covariant derivative on $M\times
\bbbc^1$ has the form 
$$
\nabla_X=X-i\frac{e}{\hbar}\alpha (X),
            \quad X\in {\cal X}(M).  $$
Two such covariant derivatives $\nabla^1$, $\nabla^2$
are isomorphic  if and only if there exists a function 
$f : M\rightarrow T^1$
such that 
$$
\alpha_2 = \alpha_1 - i\frac{\hbar}{e}
         \frac{df}{f}.$$

Following the terminology of \cite{8,12}, we say that the 
1--forms $(e/\hbar)\alpha_j$, $j=1,2$ are 
{\it logarithmically cohomologous;}
$\lambda = -i(df)/f$ is said to be {\it logarithmically
exact.} A 1--form $\lambda$ is logarithmically exact if 
and only if
$$
\exp(i\int_{\gamma}^{}\lambda )=1,\quad \mbox{i.e.}\quad 
\int_{\gamma}^{}\lambda=2\pi n,\; n\in \bbbz,$$ 
holds for all 1--cycles $\gamma$ of the singular
homology. If $\gamma$ is a periodic cycle, i.e. $p\gamma =
\partial \sigma$ for some $p\in \bbbz$ and 2-cycle $\sigma$,
then
$$
0=\int_{\sigma}^{}d\lambda= \int_{p\gamma}^{}\lambda
          =p\int_{\gamma}^{}\lambda,$$ 
because a logarithmically exact form is closed. 

So it suffices to check independent, non--periodic 1--cycles.
Their number is $b_1(M)$, the first Betti number of manifold
$M$. Now, quantum Borel kinematics on $M$ in the case 
of trivial fibration and of vanishing magnetic field 
is determined by a closed 1-form $\alpha$. 
Its cohomology class is in turn --- according to the 
de Rham Theorem --- determined by $b_1(M)$ periods
$$
\Phi_j=\int_{\gamma_j}^{}\alpha .$$ 
For each $j=1,\ldots,b_1(M)$, $\Phi_j$ represents an external
magnetic flux outside the manifold $M$ and passing through the
$j$-th independent cycle $\gamma_j$. 
Two potentials $\alpha_1$, $\alpha_2$ determine
the same quantum Borel kinematics if
$$
\frac{e}{\hbar}(\Phi_j^{(1)}-\Phi_j^{(2)}) \in 2\pi \bbbz
\quad \mbox{for all} \quad j.$$ 
We conclude that in the considered special case the family
 of all inequivalent quantum Borel kinematics can be parametrized by elements 
$$
z=(z_1,\ldots,z_{b_1(M)})\in \rm{U}(1) \times \ldots
\times \rm{U}(1), \quad z_j=\exp(ie\Phi_j/\hbar).$$

\subsection{Examples} 
In the examples we concentrate on elementary quantum Borel 
 kinematics.

\noindent {\bf Example 5.1.} 
$M=\bbbr^3$. Since $H^2(\bbbr^3, \bbbz)=0$, only  the trivial principal
bundle exists over $M$. Every smooth function  $f : \bbbr^3
\rightarrow $ U(1) can be written in the form  
$f=\exp(ie\lambda /\hbar)$ with $\lambda$ being a smooth 
real function. Two 1--forms are
logarithmically cohomologous if and only if they are
cohomologous. Thus (an equivalence class of) quantum kinematics
is obtained by taking (a cohomology class of) a closed  
2--form  $\beta$ of magnetic field. 
There exists exactly one equivalence class of quantum 
Borel kinematics (i.e. with vanishing magnetic field)
--- the standard one  with $\alpha=0$. 

\noindent {\bf Example 5.2.}
$M=\bbbr^3\backslash \bbbr$ is the 3-dimensional real space $\bbbr^3$ with
the  $x_3$\/-axis excluded (Aharonov-Bohm configuration). Then
$H^2(M,\bbbz)=0$, and  there exists only the trivial principal
bundle over $M$. There exists exactly  one (up to homology)
independent cycle in $M$ and it is non-periodic, i.e.
$H_1(M,\bbbz)=\bbbz, H^1(M,\bbbr)=\bbbr$. 
Inequivalent quantum Borel kinematics
are labeled  by $z\in $ U(1), $z=\exp(ie\Phi /\hbar)$; 
$\Phi$ denotes the magnetic flux  supported by the excluded
line $\bbbr$. The corresponding vector potential 
1--form $\alpha$ can  be chosen in the form
$$
\alpha=\frac{\Phi}{2\pi}
(-\frac{x_2}{x_1^2+x_2^2}dx_1+\frac{x_1}{x_1^2+
x_2^2}dx_2).$$ 
From the point of view of quantum mechanics a
charged particle cannot  distinguish between the flux 
$\Phi=2\pi n\hbar/e$, $n\in \bbbz$, and the zero one. 
In fact, this is the effect discovered by Y.Aharonov and 
D.Bohm \cite{1}. Let us note that the Aharonov--Bohm
effect in the presence of two solenoids or $n$ solenoids
placed along a straight line was studied in \cite{St94}.

\noindent{\bf Example 5.3.}
 $M$ be a compact orientable surface. It is known
that, up to diffeomorphism, such $M$ are classified by the
 Euler characteristic $\chi(M)=2-2p$, where $p=0,1,2,\ldots$
 is the {\it genus} of $M$. For given $p$, we denote
 $M=K_p$; it is modeled by a  2-sphere with $p$ handles
 (\cite{22}, Chap. 9.3). Since
$H^2(K_p,\bbbz)=\bbbz,$ there exist countably many principal
 bundles over $K_p,$ labeled by $n\in \bbbz.$ According to
 our physical interpretation, $n$ can by related to the
 magnetic charge of  a Dirac monopole enclosed by the closed
 surface, $g=(2\pi\hbar/e)n.$ Further, 
$H_1(K_p,\bbbz)=\bbbz^{2p};$ thus
inequivalent quantum Borel kinematics (with vanishing
 magnetic field) are labeled by elements of the dual group
 $\mbox{U}(1)^{2p}.$
Each quantum Borel kinematics depends on $2p$ external
 magnetic fluxes, each handle carrying two of them.
On a 2--torus, generalized quantum kinematics
was studied in \cite{DTtorus}, quantum Borel kinematics
in \cite{CS}.

\noindent{\bf Example 5.4.}
$M=\bbbr P^n=S^n/\{\pm\}$ is a real projective space, 
$n > 2.$ We can identify: 
$$
S\in {\cal B}(\bbbr P^n)\leftrightarrow S\in 
{\cal B}(S^n), \qquad S=-S;  
X\in {\cal X}(\bbbr P^n)\leftrightarrow X\in 
{\cal B}(S^n), \qquad X_{-u}=-X_u;   $$
$$H^2(\bbbr P^n,\bbbz)=\bbbz_2, \qquad
 H_1(\bbbr P^n),\bbbz)=\bbbz_2.
$$ 
Hence there exist two inequivalent principal bundles over
 $M$ and two inequivalent QBK's in mutually inequivalent
 fibrations. The QBK's can be explicitly described in the
 following way: the Hilbert spaces
$\cal H_+,\, H_-$ are chosen as subspaces in $L^2(S^n,d\mu)$
 (with measure $\mu$ invariant under the transformation 
$u \rightarrow -u$), $\psi\in{\cal H}_{\pm}$ if and only if
$\psi(-u)=\pm\psi(u);$ the two inequivalent systems of
imprimitivity ($c=0$) are defined for both signs + and --
 by the operators
$$
E(S)=\chi_S.,\; P(X)=-i\hbar(X+\frac{1}{2} 
  \mbox{div}_{\mu}X),\quad
S\in{\cal B}(\bbbr P^n), \, X\in{\cal X}(\bbbr P^n))$$
acting in ${\cal H}_+$ and ${\cal H}_-$, respectively. 
The real projective space $\bbbr P^n$
appears in quantum mechanics e.g. as a (topologically
non--trivial) part of the effective configuration space 
of two indistinguishable point--like particles localized 
in the $(n+1)$--dimensional Euclidean space $\bbbr^{n+1}.$
The two cases with signs + and -- correspond in quantum
mechanics to the cases of bosonic and fermionic statistics,
 respectively.
More details can be found in \cite{6}; see also \cite{9, 10}.
It should be stressed that the case $n=2$ presents 
unexpected features: in \cite{10} it was found for a system of
two particles in two dimensions that there is a continuous 
family of quantizations describing new statistics which
interpolate between fermions and bosons. 
These anomalous or fractional statistics were later discovered 
independently by \cite{GMS80, GMS81} and by \cite{W82a,W82b}, 
who actually coined the term `anyons' for the corresponding
particles.

\vskip 1cm
 
\begin{table}[t]
\caption{Examples of elementary quantum Borel kinematics.
\label{tab}}
\vspace{0.2cm}
\begin{center}
\footnotesize
\begin{tabular}{|l|c|c|c|c|c|}
\hline
{Quantum system} & \raisebox{0pt}[13pt][7pt]{$M$} &
\raisebox{0pt}[13pt][7pt]{$\pi_{1}(M)$} &
{$H_{1}(M,Z)$} & {$H^{2}(M,Z)$}&
\begin{minipage}{1in}
\begin{center}
Topological \\ quantum \\ numbers
\end{center}
\end{minipage} 
\\[5pt] 
\hline
Spinless particle in $R^{3}$   & $R^{3}$ & $\{e\}$ &
$0$  & $0$ & ---  \\[5pt]
Aharonov--Bohm configuration & $R^{3} \backslash R $ &
$Z$ & $Z$ & $0$ & $\vartheta \in [0,1)$ \\[5pt]
Dirac's monopole & $R^{3} \backslash O = R_{+} \times 
S^{2}$ & $\{e\}$ & $0$ & $Z$ & $n \in Z $  \\[5pt]
\begin{minipage}{1.5in}
2 distinguishable \\ particles in $R^{3}$
\end{minipage} 
& $R^{3} \times R_{+} \times S^{2}$ & $\{e\}$  & $0$
& $Z$ & $n \in Z $  
\\[10pt]
\begin{minipage}{1.5in}
2 indistinguishable \\ particles in $R^{3}$
\end{minipage} 
& $R^{3} \times R_{+} \times RP^{2}$ & $S_{2}$  & $Z_{2}$
& $Z_{2}$ & $m \in Z_{2} $  
\\[10pt]
Rigid body & $R^{3} \times SO(3)$ & $Z_{2}$  & $Z_{2}$
&$Z_{2}$ & $m \in Z_{2} $  
\\[5pt]
Symmetric top & $S_{2}$ & $\{e\}$ & $0$ & $Z$ & $n \in Z $
\\[5pt]
Rotator with fixed axis & $S^{1}$ & $Z$ & $Z$ & $0$
& $\vartheta \in [0,1) $
\\[5pt]
\begin{minipage}{1.5in}
Particle on orientable \\ surface of genus $p$
\end{minipage} 
& $K_{p}$ & $\pi_{1}(K_{p})$ & $Z^{2p}$ & $Z$ & 
\begin{minipage}{1in}
\begin{center}
$n \in Z, $ \\ $\vartheta_{1} \ldots \vartheta_{2p} 
\in [0,1)$
\end{center}
\end{minipage} 
\\[10pt]
\hline
\end{tabular}
\end{center}
\end{table}


\noindent ACKNOWLEDGEMENTS\\[1ex]
J. T. and P. \v{S}. are grateful to Prof. Dr. H.D. Doebner
for the kind hospitality extended to them on various
occasions at the Arnold Sommerfeld Institute in Clausthal. The authors thank A. B\'ona for typing
the manuscript. Partial support of the Grant Agency
of Czech Republic (contract No. 202/96/0218) is
acknowledged. The list of references is by no means complete
and we apologize to the authors of papers which have not
been included.


\begin{thebibliography}{99} 
\bibitem{1} Y. Aharonov, D. Bohm:
``Significance of electromagnetic potentials in the quantum
  theory'', 
  Phys. Rev. {\bf 115}, 485--491 (1959). 
\bibitem{2} B. Angermann: 
``Global geometry and Schr\"odinger quantization on
 manifolds'', 
  \em Latin American Summer School of Physics at
  UNAM, \em Mexico (1980). 
\bibitem{3} B. Angermann, H.-D. Doebner:
``Homotopy groups and quantization of localizable systems'',
  Proc. Xth Coll. Group Theoretical Methods in Physics,
  Physica {\bf 114}A, 433--439 (1982). 
\bibitem{4} B. Angermann, H.-D. Doebner, J. Tolar: 
  ``Quantum kinematics on smooth manifolds''. 
  In: \em Non-linear Partial Differential Operators and
  Quantization Procedures,\em 
  Lecture Notes in Mathematics, Vol. 1037,
  Springer-Verlag, Berlin (1983), pp. 171--208.
\bibitem{BFFLS} F. Bayen, M. Flato, C. Fronsdal,
  A. Lichn\'erowicz, D. Sternheimer: 
  ``Deformation theory and quantization'',
  Ann. Phys. (New York) {\bf 111}, 61--110, 111--152 (1978).
\bibitem{5} P.A.M. Dirac: 
  ``Quantized singularities in the electromagnetic field'',
  Proc. Roy. Soc. London A {\bf 133}, 60--72 (1931).
\bibitem{DEH} H.-D. Doebner, H.J. Elmers, W.F. Heidenreich:
  ``On topological effects in quantum mechanics; 
  The harmonic oscillator in the pointed plane'', 
  J. Math. Phys. {\bf 30}, 1053--1059 (1989). 
\bibitem{DG} H.-D. Doebner, G.A. Goldin:
  ``On a general nonlinear Schr\"odinger equation
  admitting diffusion currents'',
  Phys. Lett. A{\bf162}, 397--401 (1992).
\bibitem{6} H.-D. Doebner, P. \v S\v tov\'\i\v cek, 
  J. Tolar:
  ``Quantizations of the system of two indistinguishable
  particles'', 
  Czech. J. Phys. B {\bf 32}, 1240--1248 (1982).
\bibitem{7} H.-D. Doebner, J. Tolar:
  ``Quantum mechanics on homogeneous spaces'', 
  J. Math. Phys. {\bf 16}, 975--984 (1975). 
\bibitem{8} H.-D. Doebner, J. Tolar: 
 ``On global properties of quantum systems''. 
  In: \em Symmetries in Science \em 
  (B. Gruber and R. S. Millman, eds.) Plenum Press,
  New York (1980), pp. 475--486. 
\bibitem{Do-To} H.-D. Doebner, J. Tolar: 
  ``Symmetry and topology of the configuration space and
  quantization''. In: \em Symmetries in Science II \em 
  (B. Gruber and R. Lenczewski, eds.) 
  Plenum, New York (1986), pp. 115--126.
\bibitem{DTtorus} H.-D. Doebner, J. Tolar:
  ``Quantum particle on a torus with an external magnetic
  field'', in: {\it Quantization and Coherent States
  Methods} (S.T. Ali, I.M. Mladenov and A. Odzijewicz,
  eds.) World Scientific, Singapore (1993),
  pp. 3--10.

\bibitem{Drees} M. Drees: ``Zur Kinematik lokalisierter 
  quantenmechanischer Systeme unter Ber\"ucksichtigung
  innerer Freiheitsgrade und \"au\ss{}erer Felder'',
  Ph.D. Thesis, Technical University, Clausthal (1992).
\bibitem{GO} P. Goddard, D.I. Olive:
``Magnetic monopoles in gauge field theories'',
  Rep. Prog. Phys. {\bf 41}, 1357--1437 (1978).
\bibitem{GMS80} G.A. Goldin, R. Menikoff, D.H. Sharp:
``Particle statistics from induced representations
  of a local current group'',
  J. Math. Phys. {\bf 21}, 650--664 (1980).
\bibitem{GMS81} G.A. Goldin, R. Menikoff, D.H. Sharp:
``Representations of a local current algebra in
  non--simply connected space and the Aharonov--Bohm
  effect'', 
  J. Math. Phys. {\bf 22}, 1664--1668 (1981).
\bibitem{31} G.A. Goldin, R. Menikoff, D.H. Sharp: 
``Induced representations of the group of diffeomorphisms 
  of $\bbbr^3$'', 
  J. Phys. A: Math. Gen. {\bf 16}, 1827--1833 (1983).
\bibitem{GP} W. Greub, H.-R. Petry:
``Minimal coupling and complex line bundles'',
  J. Math. Phys. {\bf 16}, 1347--1351 (1975).
\bibitem{Ko} B. Kostant: 
``Quantization and unitary representations: Part I.
  Prequantization''. 
  Lecture Notes in Mathematics, Vol. 170,
  Springer-Verlag, Berlin (1970), pp. 87--208. 
\bibitem{9} M.G.G. Laidlaw, C. Morette-DeWitt: 
``Feynman functional integrals for
  systems of indistinguishable particles'',
  Phys. Rev. D {\bf 3}, 1375--1378 (1970). 
\bibitem{10} J.M. Leinaas, J. Myrheim: 
``On the theory of identical particles'', 
  Nuovo Cimento B {\bf 37}, 1--23 (1977). 
\bibitem{11} G.W. Mackey: 
``Unitary representations of group extensions I'', 
  Acta Math. {\bf 99}, 265--311 (1958).
\bibitem{Mi} J. Milnor: ``On the existence of a connection
  with curvature zero'', 
  Comment. Math. Helv. {\bf 32}, 215--223 (1958).
\bibitem{Mu} U.A. M\"{u}ller, H.D. Doebner: 
``Borel quantum kinematics of rank $k$ on smooth
  manifolds'',
  J. Phys. A: Math. Gen. {\bf 26}, 719--730 (1993). 
\bibitem{Natter} P. Nattermann: ``Dynamics in Borel
  quantization: Nonlinear Schr\"odinger equations
  vs. master equations'',
  Ph.D. Thesis, Technical University, Clausthal (1997).
\bibitem{12} I.E. Segal:
``Quantization of nonlinear systems'', 
  J. Math. Phys. {\bf 1}, 468--488 (1960). 
\bibitem{13} L.S. Schulman: ``A path integral for spin'',
  Phys. Rev. {\bf 176}, 1558--1569 (1968). 
\bibitem{CS} C. Schulte:
``Quantum mechanics on the torus, Klein bottle and
  projective sphere'', in: {\it Symmetries in Science IX}
  (B. Gruber ed.) Plenum Press, New York (1997), 
  pp. 313--323.
\bibitem{15} P. \v S\v tov\'{\i}\v cek: 
``Dirac monopole derived from representation theory'', 
  Suppl. Rend. Circ. Mat. Palermo, Ser. II, No. 3, 301--306
  (1984). 
\bibitem{St88} P. \v S\v tov\'{\i}\v cek:
``Systems of imprimitivity for the group of 
  diffeomorphisms  I, II'',  
  Ann. Global Anal. Geom. {\bf 5}, 89--95
  (1987), {\bf 6}, 31--37 (1988). 
\bibitem{St94} P. \v S\v tov\'{\i}\v cek: 
``Scattering on a finite chain of vortices'',
  Duke Math. J. {\bf 76}, 303--332 (1994). 
\bibitem{14} P. \v S\v tov\'{\i}\v cek, J. Tolar:
``Topology of the configuration manifold and quantum
  mechanics'',
  Acta Polytechnica (Prague) Ser. IV, No. 1 (1984),
  37--75  (in Czech). 
\bibitem{W82a} F. Wilczek:
``Magnetic flux, angular momentum and statistics'',
  Phys. Rev. Lett. {\bf 48}, 1144--1146 (1982).
\bibitem{W82b} F. Wilczek:
``Quantum mechanics of fractional spin particles'',
  Phys. Rev. Lett. {\bf 49}, 957--959 (1982).
\bibitem{16} T.T. Wu, C.N. Yang:
``Concept of nonintegrable phase factors and global
  formulation of gauge fields'', 
  Phys. Rev. D{\bf 12}, 3845--3857 (1975).
\bibitem{17} C.N. Yang: ``Magnetic monopoles, fiber bundles,
  and gauge fields'', 
  Annals of the New York Academy of Science 
  {\bf 294}, 86--97 (1977). 
\bibitem{18} N. Bourbaki: \em Groupes et alg\`ebres de Lie,
\em Hermann, Paris (1968). 
\bibitem{19} C. Chevalley: \em Theory of Lie Groups I,\em
  Princeton University Press, Princeton, N. J. (1946).
\bibitem{Davies} E.B. Davies: {\it Theory of Open Quantum
  Systems,} Academic Press, London (1976).
\bibitem{Dir} P.A.M. Dirac: {\it Lectures on Quantum
  Mechanics,} Belfer Graduate School of Science,
  Yeshiva University, New York (1964).
\bibitem{Dirac} P.A.M. Dirac: {\it The Principles of Quantum
    Mechanics}, 4th ed., Oxford University Press, Oxford
   (1958).
\bibitem{GHV} W. Greub, S. Halperin, R. Vanstone:
   \em Connections, Curvature and Cohomology, I -- III, \em
   Academic Press, New York (1973).
\bibitem{20} V. Guillemin, A. Pollack: 
  \em Differential Topology, \em  
  Prentice-Hall, Inc., Englewood Cliffs, N. J. (1974). 
\bibitem{21} S. Helgason:
  \em Differential Geometry, Lie Groups,
  and Symmetric Spaces, \em Academic Press, New York (1978).
\bibitem{22} M.W. Hirsch: \em Differential Topology, \em 
  Springer-Verlag, Berlin (1970).
\bibitem{23} A.A. Kirillov: 
  \em Elements of Representation Theory, \em 
  Springer--Verlag, Berlin (1984).
\bibitem{24} S. Kobayashi, K. Nomizu:
  \em Foundations of Differential Geometry I,\em
  Interscience--Wiley, New York (1963). 
\bibitem{25} S. Lang: 
  \em Introduction to Differentiable Manifolds, \em
  Interscience Publishers, New York (1962). 
\bibitem{26} G.W. Mackey: 
  \em Induced Representations and Quantum
   Mechanics, \em 
  W.A. Benjamin, Inc., New York (1968). 
\bibitem{27} K. Nomizu: 
  \em Lie Groups and Differential Geometry,\em 
  The Mathematical Society of Japan (1956). 
\bibitem{28} W. Pauli: \em Wellenmechanik, \em
  Handbuch der Physik Bd. 24, Teil 1, 120 (1933).
\bibitem{29} V.S. Varadarajan:
  \em Geometry of Quantum Theory I, \em
  Van Nostrand, Princeton, N. J. (1968).
\bibitem{30} V.S. Varadarajan:
  \em Geometry of Quantum Theory
  II. Quantum Theory of Covariant Systems, \em
  Van Nostrand Reinhold Co., New York (1970).
\end{thebibliography}
\end{document}